\documentclass[aps,pra,superscriptaddress,notitlepage, amsfonts,longbibliography, twocolumn]{revtex4-2}

\usepackage{qcircuit}
\usepackage[dvipdfmx]{graphicx}
\usepackage{amsmath,amssymb,amsthm,mathrsfs,amsfonts,dsfont}
\usepackage{subfigure, epsfig}
\usepackage{braket}
\usepackage{bm}
\usepackage{enumerate}
\usepackage{physics}
\usepackage{url}
\usepackage{booktabs}

\usepackage[colorlinks,linkcolor=blue,citecolor=blue]{hyperref}
\usepackage{stmaryrd}
\usepackage{mathtools}
\usepackage{comment}
\usepackage[linesnumbered,ruled,vlined]{algorithm2e}
\SetKwInOut{Input}{Input}
\SetKwInOut{Output}{Output}
 \usepackage[normalem]{ulem}
\makeatletter
\newcommand{\NoNumber}{\renewcommand{\algocf@printnl}[1]{}}
\newcommand{\ReNumber}{\renewcommand{\algocf@printnl}[1]{\algocf@nl{##1}}}
\makeatother

\makeatletter
\renewcommand*{\numberline}[1]{\hb@xt@1em{#1\hfil}} 
\makeatother

\graphicspath{{./fig/}}

\begin{document}
\title{Degeneracy Cutting: A Local and Efficient Post-Processing for Belief Propagation Decoding of Quantum Low-Density Parity-Check Codes}

\author{Kento Tsubouchi}
\email{tsubouchi@noneq.t.u-tokyo.ac.jp}
\affiliation{Nanofiber Quantum Technologies, Inc., 1-22-3 Nishiwaseda, Shinjuku-ku, Tokyo 169-0051, Japan}

\author{Hayata Yamasaki}
\email{hayata.yamasaki@nano-qt.com}
\affiliation{Nanofiber Quantum Technologies, Inc., 1-22-3 Nishiwaseda, Shinjuku-ku, Tokyo 169-0051, Japan}
\affiliation{Department of Computer Science, Graduate School of Information Science and Technology, The University of Tokyo, 7-3-1 Hongo, Bunkyo-ku, Tokyo, 113-8656, Japan}

\author{Shiro Tamiya}
\email{shiro.tamiya@nano-qt.com}
\affiliation{Nanofiber Quantum Technologies, Inc., 1-22-3 Nishiwaseda, Shinjuku-ku, Tokyo 169-0051, Japan}

\begin{abstract}
Quantum low-density parity-check (qLDPC) codes are promising for realizing scalable fault-tolerant quantum computation due to their potential for low-overhead protocols.
A common approach to decoding qLDPC codes is to use the belief propagation (BP) decoder, followed by a post-processing step to enhance decoding accuracy.
For fast decoding, the post-processing algorithm is desirable to have a small computational cost and rely only on local operations on the Tanner graph to facilitate parallel implementation.
To address this requirement, we propose \textit{degeneracy cutting} (DC), an efficient post-processing technique for the BP decoder that operates on information restricted to the support of each stabilizer generator.
DC selectively removes one variable node with the lowest error probability for each stabilizer generator, significantly improving decoding performance while retaining the favorable computational scaling and structure amenable to parallelization inherent to BP.
We further extend our method to realistic noise models, including phenomenological and circuit-level noise models, by introducing the \textit{detector degeneracy matrix}, which generalizes the notion of stabilizer-induced degeneracy to these settings.
Numerical simulations demonstrate that BP+DC achieves decoding performance approaching that of BP followed by ordered statistics decoding (BP+OSD) in several settings, while requiring significantly less computational cost.
Our results present BP+DC as a promising decoder for fault-tolerant quantum computing, offering a valuable balance of accuracy, efficiency, and suitability for parallel implementation.
\end{abstract}
\maketitle

\section{Introduction}
Fault-tolerant quantum computation is essential for realizing quantum computation with noisy quantum devices~\cite{shor1996fault, aharonov1997fault, 10.5555/2011665.2011666, gottesman2009introduction, nielsen2010quantum, gottesman2013fault,tamiya2024polylog, Yamasaki_2024}.
Quantum low-density parity-check (qLDPC) codes~\cite{Tillich_2014, Leverrier_2015, Hastings_2021, breuckmann2021quantum, Panteleev_2022, 10.1145/3519935.3520017, 10.1145/3564246.3585101, bravyi2024high} are promising candidates for implementing scalable fault-tolerant quantum computation due to their potential to enable low resource overhead~\cite{gottesman2013fault, fawzi2020constant, cohen2022low, tamiya2024polylog, xu2024constant, yoder2025tour, nguyen2025quantum}.
A key challenge to enable scalable fault-tolerant quantum computation with qLDPC codes is the development of a decoder that is both fast and accurate for large-scale computation~\cite{Terhal_2015, Barber_2025, takada2025doubly}.

\begin{figure*}[t]
    \begin{center}
        \includegraphics[width=0.8\linewidth]{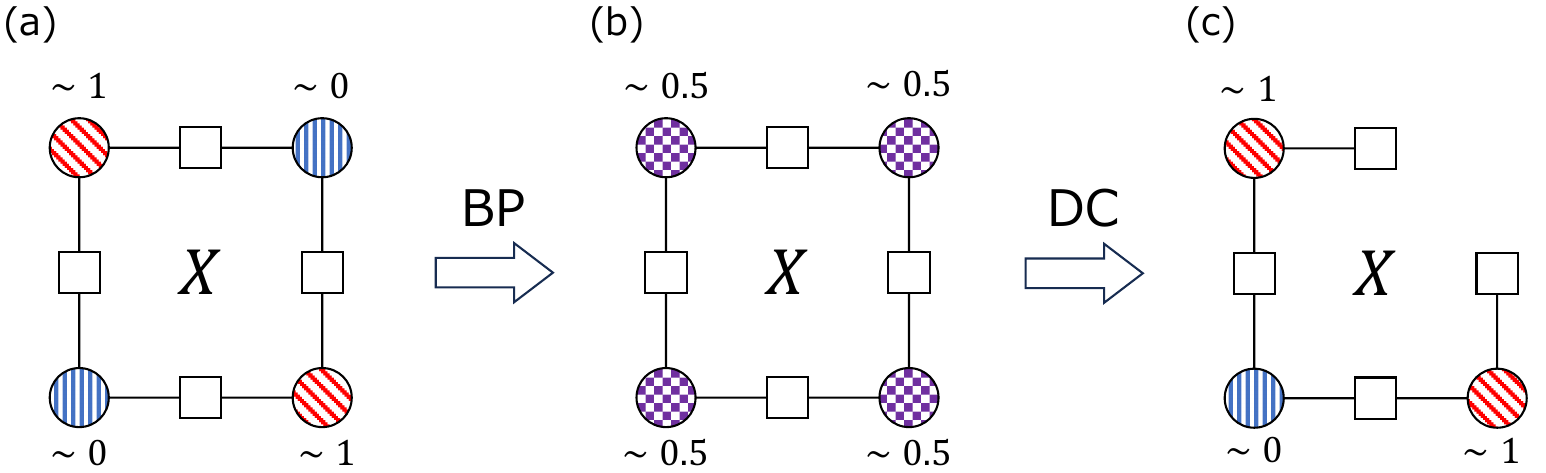}
        \caption{
        An illustration of degeneracy in belief propagation (BP) decoding and degeneracy cutting (DC).
        Circles represent variable nodes (qubits) in the support of an $X$-type stabilizer generator, while squares represent check nodes (parity checks) connected to the variable nodes.
        (a) A physical error pattern affecting two qubits (circles with red diagonal patterns).
        (b) BP cannot distinguish this pattern from an alternative two-qubit error (circles with blue vertical patterns shown in panel (a)) that produces the identical syndrome.
        This ambiguity arises from the degeneracy of the code.
        Consequently, the decoder assigns nearly equal estimated error probabilities to all four involved qubits (circles with purple checkerboard patterns), leading to a decoding failure.
        While the posterior probabilities may either remain close to $\sim0.5$~\cite{roffe2020decoding, yin2024symbreak} or oscillate between $0$ and $1$~\cite{chytas2025enhanced}, this figure represents the case where all estimated error probabilities are near $0.5$.
        (c) The variable node with the lowest error probability is identified for each stabilizer generator and removed from the Tanner graph.
        If several variable nodes have the same lowest error probability, the nodes to be removed are selected at random.
        This process breaks the degeneracy introduced by the stabilizers, enabling an accurate error estimate to be obtained by rerunning BP.
        }
        \label{fig_degeneracy_cutting}
    \end{center}
\end{figure*}

The belief propagation (BP) decoder is widely and successfully used for decoding classical LDPC codes due to its excellent performance and linear-time complexity~\cite{Gallager1962,mackay1996near,richardson2001design,kschischang2001factor}.
When BP is applied to a qLDPC code, the algorithm operates on the Tanner graph, a bipartite graph whose variable nodes represent physical qubits and whose check nodes represent stabilizer generators~\cite{1056404,poulin2008iterative}.
However, their performance for qLDPC codes is significantly degraded by the degeneracy of quantum codes~\cite{raveendran2021trapping,fuentes2021degeneracy,poulin2008iterative}, where the decoder cannot distinguish between multiple error patterns that yield the same syndrome, often resulting in decoding failure (see Fig.~\ref{fig_degeneracy_cutting} (b)).
To address this issue, a post-processing technique called ordered statistics decoding (OSD) has been proposed~\cite{panteleev2021degenerate, roffe2020decoding}.
While standard BP decoding scales linearly with code size, the OSD step incurs a computational cost of $O(n^3)$ by requiring solving a system of linear equations (typically via matrix inversion or Gaussian elimination), where $n$ is the number of physical qubits of a qLDPC code~\cite{panteleev2021degenerate, roffe2020decoding}, posing a practical obstacle for fast decoding.
To reduce the decoding time further, several alternative post-processing methods have been proposed~\cite{du2022stabilizer,du2024check,yao2024belief, gong2024toward,iolius2024closed,hillmann2024localized,wolanski2024ambiguity,iolius2024almost,yin2024symbreak,ott2025decision}.
While some of these techniques achieve linear time complexity comparable to BP, they often rely on global information extracted from the BP output, such as error probabilities compared across all variable nodes in the Tanner graph.
For fast decoding, however, it is desirable to have a computationally efficient decoder that makes decisions based only on local information restricted to the support of individual stabilizer generators.
This locality is necessary for parallel execution, which is crucial for achieving fast decoding.

In this work, we propose \textit{degeneracy cutting} (DC), an efficient post-processing technique for BP decoding that operates on information restricted to the support of each stabilizer generator.
The DC algorithm proceeds by identifying, for each stabilizer generator, the variable node with the lowest error probability as estimated by BP, and then removing that node from the Tanner graph before performing a second round of BP (see Fig.~\ref{fig_degeneracy_cutting} (c)).
Degeneracy in a CSS code occurs whenever two error patterns differ by an element of the stabilizer yet yield the same syndrome; by pruning the lowest-probability qubit within the support of each stabilizer generator, DC eliminates precisely these locally degenerate alternatives.
In qLDPC codes, each stabilizer generator acts only on a constant number of qubits, ensuring the computational cost of DC scales linearly with the code size.
Furthermore, unlike previous linear-time post-processing methods~\cite{iolius2024closed, iolius2024almost, yin2024symbreak}, which compare marginal probabilities across all variable nodes, DC performs comparisons only among variable nodes associated with the same stabilizer generator. As a result, the decisions made by DC do not rely on global comparisons, making the algorithm amenable to parallel implementation.

We evaluate the performance of the BP+DC decoder for surface codes~\cite{dennis2002topological, kitaev2003fault, fowler2012surface} and bivariate bicycle (BB) codes~\cite{bravyi2024high} under the code-capacity noise model, in which syndrome extraction is assumed to be noiseless while only data qubits are subject to errors. 
Our numerical simulations show that BP+DC achieves a decoding accuracy approaching that of the BP+OSD decoder, and for BB codes, it exhibits a lower decoding failure probability than BP+OSD, while offering a reduction in computational cost from $O(n^3)$ to $O(n)$.

We also extend our approach to more realistic phenomenological and circuit-level noise models, which both incorporate errors into the syndrome extraction process~\cite{dennis2002topological, fowler2012surface}.
In these settings, degeneracy may arise from measurement errors or from errors propagating through auxiliary qubits that are used for syndrome extraction, further degrading the performance of BP decoding.
To address these new sources of degeneracy, we introduce a \textit{detector degeneracy matrix}, which generalizes the notion of trivial errors beyond those characterized by the stabilizer generators.
This matrix includes not only the original stabilizer generators but also newly identified circuit-level degeneracies.
When applying DC in these noise models, we use the detector degeneracy matrix instead of the stabilizer generators. 
This approach enables the direct application of stabilizer-based post-processing techniques~\cite{du2022stabilizer, yin2024symbreak}, previously limited to the code-capacity noise model, to more realistic noise models.
By constructing the detector degeneracy matrix with identified circuit-level trivial-error patterns up to weight $w=5$, we numerically demonstrate that the BP+DC decoder maintains a performance comparable to BP+OSD even in these more realistic settings.

Our framework is inherently extensible, and we expect that including higher-weight trivial errors is a promising direction for further enhancement.

The results of this work address the computational constraints of decoder, where substantial overhead is expected from the need for a high-bandwidth classical--quantum interface.
In such settings, a favorable decoding strategy is determined not only by its overall scaling and performance, but also by its simplicity and parallelizability.
A decoder such as BP+DC, which avoids global information exchange and is structured for parallel processing, therefore provides a preferable option for hardware-conscious implementations.
By reducing the classical computation overhead associated with decoding, our method contributes to the development of simpler and more efficient classical decoding architectures for fault-tolerant quantum computation~\cite{sunami2025transversalsurfacecodegamepowered,PRXQuantum.6.010101,Barber_2025,Terhal_2015}.

The remainder of this paper is organized as follows. In Sec.~\ref{sec_preliminaries}, we review the basics of quantum error correction, focusing on qLDPC codes and their decoders.
In Sec.~\ref{sec_degeneracy_cutting}, we introduce our proposed degeneracy cutting (DC) algorithm and evaluate its performance under the code-capacity noise model.
In Sec.~\ref{sec_beyond}, we extend our method to more realistic noise models beyond the code-capacity noise model, including phenomenological and circuit-level noise models. 
Finally, we summarize our contributions and discuss future directions in Sec.~\ref{sec_conclusion}.

\section{Preliminaries}
\label{sec_preliminaries}
In this section, we introduce the basics of quantum error correction, focusing mainly on qLDPC codes and their decoders.
We begin with an overview of stabilizer codes (Sec.~\ref{sec_preliminaries_stab}), followed by a description of qLDPC codes (Sec.~\ref{sec_preliminaries_qldpc}).
We then formulate the decoding problem for qLDPC codes and explain the belief propagation (BP) decoder along with several post-processing methods (Sec.~\ref{sec_preliminaries_decoders}).
Note that this section and Sec.~\ref{sec_degeneracy_cutting} assume the code-capacity noise model, where errors occur only on data qubits and syndrome measurements are noiseless. 
We extend our discussion to more general noise models in Sec.~\ref{sec_beyond}.

\subsection{Stabilizer codes}
\label{sec_preliminaries_stab}
Let $\mathbb{F}_2$ be the finite field with elements $\{0,1\}$ and $\mathbb{F}_2^n$ be the $n$-dimensional row vector space over $\mathbb{F}_2$.
Let $\{\ket{0},\ket{1}\}$ be the computational basis for the single-qubit Hilbert space $\mathbb{C}^2$.
Consider an $n$-qubit Hilbert space $\qty(\mathbb{C}^2)^{\otimes n}$.
Let $X$, $Y$, and $Z$ be the single-qubit Pauli operators defined by $
X \coloneqq \ket{0}\bra{1}+\ket{1}\bra{0},
Y \coloneqq \mathrm{i}\ket{1}\bra{0}-\mathrm{i}\ket{0}\bra{1}$, and $
Z \coloneqq \ket{0}\bra{0}-\ket{1}\bra{1}$, where $\mathrm{i}\coloneqq\sqrt{-1}$.
Let $I\coloneqq\ket{0}\bra{0}+\ket{1}\bra{1}$ denote the identity operator.
The Pauli group on $n$ qubits, denoted by $\mathcal{P}_n$, consists of matrices of the form $P=\alpha\bigotimes_{i=1}^n P_i$, where $\alpha\in\{\pm1, \pm \mathrm{i}\}$, $P_i\in\{I, X, Y,Z\}$.

A \textit{stabilizer} $\mathcal{S}$ is an Abelian subgroup of $\mathcal{P}_n$ that does not contain $-I^{\otimes n}$.
A \textit{stabilizer code} $\mathcal{C}$ is a linear subspace of the $n$-qubit Hilbert space defined as the common $+1$-eigenspace for all elements in $\mathcal{S}$, i.e., $\mathcal{C}\coloneqq \{\ket{\psi}\mid s\ket{\psi}=\ket{\psi}~\text{for all~$s\in\mathcal{S}$}\}$.
If $\mathcal{S}$ is generated by $n-k$ independent elements (called \textit{stabilizer generators}), the stabilizer code $\mathcal{C}$ has dimension $2^k$; we say the code $\mathcal{C}$ encodes $k$ logical qubits.
Let $\mathcal{N}(\mathcal{S})\coloneqq \{P\in\mathcal{P}_n\mid P\mathcal{S}P^{-1}=\mathcal{S}\}$ be the normalizer of $\mathcal{S}$, where $P\mathcal{S}P^{-1}\coloneqq \{PsP^{-1}\mid s\in\mathcal{S}\}$.
Let $\mathcal{Z}\coloneqq \{\alpha I^{\otimes n}\mid\alpha\in\{\pm1, \pm \mathrm{i}\}\}$ be the center of $\mathcal{P}_n$, which is the set of elements that commute with all elements in $\mathcal{P}_n$.
The logical Pauli group is the quotient group $\mathcal{L}\coloneqq \mathcal{N}(\mathcal{S})/(\mathcal{S}\cdot \mathcal{Z})$, where $\mathcal{S}\cdot \mathcal{Z}=\{sz\mid s\in\mathcal{S}, z\in\mathcal{Z}\}$.
An operator $L\in\mathcal{N}(\mathcal{S})$ is called a \textit{logical operator}. It is called \textit{nontrivial} if its coset $[L]\in\mathcal{L}$ is nontrivial, i.e., $[L]\neq [I]$.
The distance $d$ of the stabilizer code is defined as $d\coloneqq \min\{|L|\mid L\in \mathcal{N}(\mathcal{S})\setminus(\mathcal{S}\cdot \mathcal{Z})\}$, where the weight $|P|$ of a Pauli operator $P=\alpha\bigotimes_{i=1}^n P_i$ is the number of physical qubits on which it acts non-trivially, i.e., $|P|\coloneqq\#\{i\in\{1,\ldots, n\}\mid P_i\neq I\}$.
A stabilizer code with physical qubits $n$, logical qubits $k$, and code distance $d$ is denoted by an $[[n, k, d]]$ code.
In this paper, \emph{data qubits} refer to the $n$ physical qubits constituting the stabilizer code.

\textit{Calderbank--Shor--Steane (CSS) codes} are a class of stabilizer codes constructed from a pair of classical linear codes~\cite{Calderbank_1996,Steane_1996}.
The stabilizer generators of a CSS code are composed entirely of tensor products of $I$ and $X$ ($X$-type generators) or $I$ and $Z$ ($Z$-type generators).
Each type of stabilizer generator can also be described by binary matrices, called parity-check matrices, $H_X\in\mathbb{F}_2^{m_X\times n}$ and $H_Z\in\mathbb{F}_2^{m_Z\times n}$, where $m_X$ ($m_Z$) is the number of $X$-type (or $Z$-type) stabilizer generators.
The $(i,j)$-th entry of $H_X$ (or $H_Z$) is $1$ if and only if the $j$-th qubit in the $i$-th $X$-type (or $Z$-type) generator is an $X$ (or $Z$) operator.
The requirement that all stabilizer generators commute is equivalent to the condition $H_ZH_X^\top=0$.
The logical $X$ and $Z$ operators of a CSS code are also of $X$-type and $Z$-type, respectively.
Their supports can be represented by \textit{logical operator matrices} $O_X \in \mathbb{F}_2^{k \times n}$ and $O_Z\in \mathbb{F}_2^{k \times n}$~\cite{ott2025decision}.
The $(i,j)$-th entry of $O_X$ (or $O_Z$) is $1$ if and only if the $j$-th qubit in the $i$-th logical $X$ (or $Z$) operator is acted upon by $X$ (or $Z$).
Since the logical operators must commute with stabilizer generators, we have the condition $O_ZH_X^{\top} = O_XH_Z^{\top} = 0$.

\subsection{qLDPC codes}
\label{sec_preliminaries_qldpc}
A \textit{quantum low-density parity-check (qLDPC) code} is a member of a family of stabilizer codes for which the number of physical qubits involved in each stabilizer generator and the number of stabilizer generators each physical qubit is involved in are both bounded by some constants independent of $n$ as $n\rightarrow\infty$.
In this work, we focus on CSS subclasses of qLDPC codes.
For CSS-type qLDPC codes, the row and column weights of the parity-check matrices $H_X$ and $H_Z$ are bounded by constants $r$ and $c$, respectively, for all $n$.
In contrast, from the definition of code distance, each row of the logical operator matrices $O_X$ and $O_Z$ must have weight at least $d$.
In what follows, we introduce the two qLDPC families: \emph{surface codes}~\cite{fowler2012surface,Bombin_2007} and \emph{bivariate bicycle (BB) codes}~\cite{bravyi2024high}, which will later serve as testbeds for evaluating our BP+DC decoder.

\subsubsection{Surface codes}
The \emph{surface code} is a planar realization of Kitaev’s toric code \cite{dennis2002topological,kitaev2003fault,fowler2012surface}.  
In this work, we consider the \emph{rotated} surface code of code distance $d$~\cite{Bombin_2007,Tomita_2014}, which is defined on a $d\times d$ square lattice and has parameters $[[d^2,1,d]]$.  
Data qubits are on vertices, auxiliary qubits for syndrome measurements are placed at the centers of the plaquettes, and both $X$- and $Z$-type stabilizer generators occupy plaquettes in an alternating checkerboard pattern~\cite{Bombin_2007,Tomita_2014}.
Each stabilizer generator acts non-trivially on at most four data qubits, giving a row weight of $r = 4$, while every data qubit participates in at most two $X$-type and two $Z$-type generators, so the column weight is $c=2$, independent of $d$.

\subsubsection{Bivariate bicycle codes}
\label{sec_bb_code}
Bivariate bicycle (BB) codes~\cite{bravyi2024high} are a class of qLDPC codes implemented on a two-layer qubit architecture.
Let $I_l \in \mathbb{F}_2^{l \times l}$ denote the identity matrix and $S_l \in \mathbb{F}_2^{l \times l}$ the cyclic shift matrix, defined by $(S_l)_{i,j} = 1$ if $j = i+1 \bmod l$, and $0$ otherwise.
The construction uses two commuting matrices $x = S_l \otimes I_m$ and $y = I_l \otimes S_m$, which satisfy $x^l = y^m = I_{lm}$.
We select six distinct monomials $A_1, A_2, A_3, B_1, B_2, B_3$, where each monomial is a power of either $x$ or $y$.
These are used to form two matrices $A = A_1 + A_2 + A_3$ and $B = B_1 + B_2 + B_3$.
The $X$- and $Z$-type parity-check matrices of a BB code are then given by $H_X = (A \mid B)$ and $H_Z = (B^{\top} \mid A^{\top})$, respectively.
Here $\top$ denotes the transpose of a matrix.
This construction yields a code with physical qubits $n = 2lm$, with checks $m_X = m_Z = lm = n/2$, row weight $r = 6$, and column weight $c = 3$.
Different BB codes can be obtained by changing the parameters $l, m$, and the choices of monomials $A_i, B_j$.
The specific BB codes used in this work are listed in Table~\ref{tab_BB}.
\begin{table}[t]
    \centering
    \begin{tabular}{c c c c}
        \toprule
        $[[n,k,d]]$ & $(l,m)$ & $A$ & $B$ \\
        \midrule
        $[[72,12,6]]$  & $(6,6)$   & $x^3 + y + y^2$ & $y^3 + x + x^2$ \\
        $[[108,8,10]]$ & $(9,6)$   & $x^3 + y + y^2$ & $y^3 + x + x^2$ \\
        $[[144,12,12]]$& $(12,6)$  & $x^3 + y + y^2$ & $y^3 + x + x^2$ \\
        \bottomrule
    \end{tabular}
    \caption{
    Parameters of the BB codes used in this work.
    The integer pair $(l,m)$ fixes the block length $n = 2l m$ through the commuting shift operators 
    $x = S_{l}\!\otimes\!I_m$ and $y = I_{l}\!\otimes\!S_m$.  
    The matrices $A$ and $B$ are a sum of three distinct powers of $x$ or $y$, and define the parity-check matrices of CSS codes as $H_X=[A\mid B]$ and $H_Z=[B^{\top}\mid A^{\top}]$.
    Code distance of these codes was analytically computed in Ref.~\cite{bravyi2024high} using the mixed integer programming approach in Ref.~\cite{landahl2011fault}.
    }
    \label{tab_BB}
\end{table}

\subsection{Decoding qLDPC codes}
\label{sec_preliminaries_decoders}
Errors occurring on physical qubits may be detected and corrected through measuring the stabilizer generators.
In CSS codes, since the generators are separated into $X$-type and $Z$-type subsets, bit-flip ($X$) and phase-flip ($Z$) errors can be treated independently: bit-flip errors are detected by $Z$-type generators, and phase-flip errors are detected by $X$-type generators.
In this work, for simplicity, we restrict our attention to bit-flip errors.

A bit-flip error can be represented as a binary row vector $\vb{e}_X \in \mathbb{F}_2^n$, where the $i$-th entry of $\vb{e}_X$ is $1$ if and only if a bit-flip error occurred on the $i$-th physical qubit.
Measuring the $Z$-type stabilizer generators (corresponding to the rows of $H_Z$) yields a binary vector called the \textit{syndrome}, $\vb{s}_Z = \vb{e}_XH_Z^\top$.
The task of \textit{decoding} is, given a syndrome $\vb{s}_Z$, to return a recovery operation $\hat{\vb{e}}_X$ that is consistent with the syndrome, i.e., $\hat{\vb{e}}_XH_Z^\top = \vb{s}_Z$.
We say that the decoding is \textit{successful} if the returned $\hat{\vb{e}}_X$ is consistent with the syndrome and the residual error $\vb{e}_{\text{res}} = \vb{e}_X + \hat{\vb{e}}_X$ is a trivial error (i.e., an $X$-type stabilizer element), which is equivalent to the condition $\vb{e}_{\text{res}} \in \mathrm{rowspace}(H_X)$, where $\mathrm{rowspace}(H_X) \coloneqq \{\vb{u} H_X \in\mathbb{F}_2^{n} \mid \vb{u}\in\mathbb{F}_2^{m_X}\}$ denotes the $\mathbb{F}_2$‑linear subspace spanned by the rows of $H_X$.
Conversely, a decoding failure occurs if either of the following events happens:
(i) the decoder returns an estimate $\hat{\vb{e}}_X$ that is consistent with the syndrome, but the residual error is not a trivial error ($\vb{e}_{\text{res}}O_Z^\top \neq \vb{0}$), resulting in a logical error.
(ii) the decoder fails to return an estimate $\hat{\vb{e}}_X$ that is consistent with the syndrome ($\hat{\vb{e}}_XH_Z^\top \neq \vb{s}_Z$).
In this work, we use the term decoding failure probability to refer to the combined probability of these two failure events.

A crucial distinction between classical and quantum codes is \textit{degeneracy}: distinct physical errors $\vb{e}$ and $\vb{e}'$ that differ by an element of stabilizer as $\vb{e}+\vb{e}'\in \mathrm{rowspace}(H_X)$ may produce the same syndrome and act identically on the encoded quantum state.
Consequently, the decoder needs only to identify the coset $\vb{e}_X + \mathrm{rowspace}(H_X)$ rather than a unique error vector; any representative in this coset constitutes a valid recovery.
In decoding CSS codes, it is important to take the effect of degeneracy into account, which especially matters in the decoding of qLDPC codes since low-weight stabilizer generators create a large number of plausible, degenerate error patterns.
Since the true error $\vb{e}_X$ is unknown, decoding typically aims to estimate the most probable coset $\hat{\vb{e}}_X + \mathrm{rowspace}(H_X)$ that reproduces the syndrome as $\hat{\vb{e}}_XH_Z^\top = \vb{s}_Z$.

\subsubsection{Belief propagation decoder}
The \textit{belief propagation (BP) decoder} is an iterative message-passing algorithm whose application to error correction was pioneered by Gallager for decoding classical LDPC codes~\cite{Gallager1962}.
Given the prior bit-flip probabilities $\vb{p} = (p_1,\ldots,p_n)$, the $Z$-type parity-check matrix $H_Z$, and an observed syndrome $\vb{s}_Z$, the decoder estimates the posterior marginal probability
\begin{equation}
\label{eq_marginal_error_probability_vector}
    \vb{\hat{p}}=\qty(\hat{p}_1,\ldots, \hat{p}_n),
\end{equation}
where
\begin{equation}
    \label{eq_marginal_error_probability}
    \hat{p}_i=\mathrm{Pr}((\vb{e}_X)_i =1 \mid \vb{s}_Z)
    = \sum_{\sim (\vb{e}_X)_i=1} \mathrm{Pr}(\vb{e}_X \mid \vb{e}_XH_Z^\top = \vb{s}_Z)
\end{equation}
with $\sum_{\sim (e_X)_i=1}$ denoting a summation over all error vectors with $i$-th element fixed to 1.
The resulting value $\hat{p}_i\in[0,1]$ is called a \emph{soft decision}: it retains the full probability that qubit $i$ has flipped, whereas a \textit{hard decision} collapses this probability to the binary outcome.
Using these marginal probabilities, the decoder then makes a hard decision on each qubit, declaring an error on qubit $i$ whenever $\hat{p}_i \ge 0.5$,
\begin{equation}
    \label{eq_hard_decision}
    (\hat{\vb{e}}_X)_i =
    \begin{cases}
        1 & \text{if } \hat{p}_i \ge 0.5,\\
        0 & \text{otherwise.}
    \end{cases}
\end{equation}

The BP decoder iteratively estimates the marginal probabilities using a \textit{Tanner graph}.
A Tanner graph is a bipartite graph consisting of two types of nodes: variable nodes $V = \{v_i\}_{i=1}^n$, corresponding to physical qubits or columns of the parity-check matrix $H_Z$, and check nodes $C = \{c_i\}_{i=1}^{m_Z}$, corresponding to $Z$-type stabilizer generators or rows of $H_Z$.
An edge connects a variable node $v_j$ and a check node $c_i$ if and only if the $(i,j)$-th entry of $H_Z$ is 1.
At each iteration in BP decoders, probability messages are exchanged along every edge of the Tanner graph~\cite{mackay1996near, kschischang2001factor, richardson2001design, roffe2020decoding}. 
First, every variable node sends a message representing its current error probability estimate to each connected check node.
Next, each check node applies the parity constraint to update these estimates and sends the revised values back along the same edges to its connected variable nodes. 
This sequence of forward and backward message exchanges defines one BP iteration.
After each iteration the algorithm updates the vector of marginal error probabilities $\vb{\hat{p}}=\qty(\hat{p}_1,\ldots,\hat{p}_n)$ defined in \eqref{eq_marginal_error_probability_vector} and forms a tentative error estimate $\vb{\hat{e}}_X$ by the hard-decision rule in \eqref{eq_hard_decision}.
If $\vb{\hat{e}}_X$ satisfies the syndrome constraint $\hat{\vb{e}}_XH_Z^\top = \vb{s}_Z$, the procedure terminates and outputs the current $\vb{\hat{e}}_X$ and $\vb{\hat{p}}$;
otherwise it continues until a preset maximum number of iterations $T_{\mathrm{iter}}$ is reached, in which case the decoder is declared to have failed.
We note that there are several variants of message-passing rules, such as the product-sum and minimum-sum methods, both of which are used in this work.
For details on these update rules, see, e.g., Ref.~\cite{poulin2008iterative} for the product-sum method and Ref.~\cite{roffe2020decoding} for the minimum-sum method.

A notable advantage of BP decoding for LDPC codes is its computational efficiency.
The computational cost of BP per iteration grows linearly with the code length $n$.
Because each message update depends on local neighborhood information on the Tanner graph, BP is well-suited for parallel hardware implementation~\cite{1255474}.

However, relying solely on the standard BP decoder often fails to decode qLDPC codes accurately because of degeneracy~\cite{poulin2008iterative, fuentes2021degeneracy, raveendran2021trapping}.
As an illustrative example as shown in Fig~\ref{fig_degeneracy_cutting} (a), consider two errors $\vb{e}_X$ or $\vb{e}_X'$ with the same weight that differ by an $X$-type stabilizer generator as $\vb{e}_X + \vb{e}_X'=\vb{h}_X$, where $\vb{h}_X$ is a row vector of $H_X$.
Since a stabilizer generator acts trivially on the encoded information, both $\vb{e}_X$ and $\vb{e}_X'$ constitute valid recovery operations.
However, in the BP decoder, the posterior marginals in~\eqref{eq_marginal_error_probability} may assign identical probabilities to the two error patterns. 
For example, the posterior probabilities in the support of $\vb{h}_X$ may all take values close to $0.5$~\cite{roffe2020decoding, yin2024symbreak}
or oscillate between $0$ and $1$~\cite{chytas2025enhanced}.
In either case, the hard decision may fail to reproduce either of the valid error patterns, leading to a decoding failure.

\subsubsection{Post-processing for BP decoder}
Although the BP decoder alone is not effective for decoding qLDPC codes, the estimated marginal error probabilities $\vb{\hat{p}}$ obtained from BP can be post-processed to produce a more accurate correction $\hat{\vb{e}}_X$.
A prominent example of such post-processing is \textit{ordered statistics decoding} (OSD): the columns of a parity-check matrix with the smallest $\hat{p}_i$ values are temporarily removed so that the reduced parity-check matrix becomes invertible; one then solves a linear equation $\vb{e}_XH_Z^\top=\vb{s}_Z$ using methods such as matrix inversion or Gaussian elimination~\cite{panteleev2021degenerate, roffe2020decoding}.
While BP+OSD significantly improves decoding accuracy, solving the linear equations can incur a computational cost that scales cubically with the number of physical qubits $n$.
For fast decoding, a lighter post-processing algorithm is desirable, ideally matching BP’s $O(n)$ computational cost and relying only on local operations on the Tanner graph so that the post-processing can be parallelized.

To this end, many alternative post-processing methods have been proposed.
Localized statistics decoding~\cite{hillmann2024localized} and ambiguity clustering~\cite{wolanski2024ambiguity} divide the Tanner graph into small, disjoint clusters, and solve a system of linear equations within each cluster.
Although this approach allows for parallel execution across clusters, the worst-case computational cost scales cubically.
By contrast, closed branch decoding~\cite{iolius2024closed} and decision-tree decoding~\cite{ott2025decision} avoid solving a system of linear equations: they sequentially assign errors to variable nodes one by one by deciding whether to flip each qubit based on the BP outputs, until the syndrome is satisfied.
While these methods typically exhibit a lower average-case computational cost than BP+OSD, their runtime can scale poorly in the worst case, namely, over certain error patterns or decoding paths that require deep search, unless a cutoff is applied.

Another class of post-processing algorithms modifies the Tanner graph and reruns BP.
For instance, stabilizer inactivation~\cite{du2022stabilizer} and SymBreak~\cite{yin2024symbreak} alter the graph according to the $X$-type stabilizer generators.
Stabilizer inactivation first identifies the $X$-type stabilizer generators whose qubits in the support have the smallest average reliability $|\log((1-\hat{p}_i)/\hat{p}_i)|$ and then removes all corresponding variable nodes in its support, so BP no longer uses them.
Instead of removing those variable nodes, Symbreak finds the check nodes that are connected to them and splits those checks into separate checks, so that the degeneracy disappears.
Check-agnosia~\cite{du2024check} and guided decimation~\cite{yao2024belief, gong2024toward} remove check nodes with the lowest reliability and variable nodes with the highest reliability, respectively.
More specifically, guided decimation modifies the Tanner graph by fixing the most reliable variable node, while check-agnosia removes the variable nodes connected to the least-reliable check nodes.
All of these methods repeatedly run BP and modify the Tanner graph as described above, based on its output $\vb{\hat{p}}$, until the syndrome condition is satisfied.
In the worst case, these procedures may be repeated up to a number of times proportional to $n$, which corresponds to the number of variable nodes, check nodes, or $X$-type stabilizer generators.
Given that the cost of each BP execution is itself proportional to $n$, the total worst-case computational cost can scale quadratically.
Although the computational cost can be reduced to $O(n)$ by imposing an upper bound on the number of Tanner graph modifications and BP reruns, they still rely on global comparisons of the output $\vb{\hat{p}}$, which undermines the locality of BP.
This issue also arises in ordered Tanner forest~\cite{iolius2024almost}.
This method removes variable nodes with small error probabilities, transforming the Tanner graph into a forest, which is a collection of disjoint trees.
While this method modifies the graph only once, it requires sorting $\vb{\hat{p}}$, which introduces a computational cost of $O(n\log n)$ and relies on global decision-making.

Moreover, in addition to noise on the data qubits, the syndrome extraction process itself is also affected by errors, as we will discuss in Sec.~\ref{sec_beyond}.
In these settings, additional degeneracy may arise from measurement errors or from error propagation through ancillary qubits used for syndrome extraction.
Therefore, for practical use, post-processing algorithms should also be capable of addressing such degeneracies.
While some methods remain effective under this noise model, stabilizer-based post-processing techniques—such as stabilizer inactivation and SymBreak—cannot be straightforwardly applied, since degeneracy no longer stems solely from $X$-type stabilizer generators.

In summary, post-processing methods discussed above do not fully match the operational simplicity of BP: they often introduce greater computational cost or rely on global comparisons of error probabilities.
For fast decoding, it is crucial to develop a post-processing method that operates solely on local decisions, maintains a computational cost of $O(n)$, and is applicable to errors occurring during the syndrome extraction process.

Separate from the post-processing methods described above, other recently proposed approaches aim to improve decoding performance by modifying the BP algorithm itself. For instance, some works incorporate code automorphisms to enhance message-passing updates~\cite{koutsioumpas2025automorphism}, introduce memory effects into BP iterations~\cite{muller2025improved}, or leverage the behavior of oscillating bits during BP to improve decoding~\cite{wang2025fully}.
The integration of these approaches, which modify the BP algorithm itself, with lightweight post-processing techniques, like our DC method introduced below, presents a promising direction for future research.

\section{Our method: degeneracy cutting}
\label{sec_degeneracy_cutting}
In this section, we present degeneracy cutting (DC), a linear-time post-processing algorithm for BP decoding that relies solely on local decisions on the Tanner graph. 
We begin with a detailed description of the algorithm (Sec.~\ref{sec_degeneracy_cutting_algorithm}).
Then, we present numerical results for surface codes and BB codes under the code-capacity noise model (Sec.~\ref{sec: Numerical analysis}).

\subsection{Description of algorithm}
\label{sec_degeneracy_cutting_algorithm}

\begin{algorithm}[t]
\caption{BP+DC decoder}
\label{alg_1}
\Input{Parity-check matrices $H_X$, $H_Z$; observed syndrome $\vb{s}_Z$; prior error probabilities $\vb{p}$; maximum number of BP iterations $T_{\mathrm{iter}}$}
\Output{Estimated $X$-type error $\hat{\vb{e}}_X$}
$(\hat{\vb{e}}_X, \hat{\vb{p}}) \leftarrow \text{BP\_decode}(H_Z, \vb{s}_Z, \vb{p}, T_{\mathrm{iter}})$\;
\If{$\hat{\vb{e}}_XH_Z^\top = \vb{s}_Z$}{
  \Return{$\hat{\vb{e}}_X$}
}
$\mathrm{Cut\_indexes} \gets \{\}$\;
\For{each row vector $\vb{h}_X$ in $H_X$}{
  $i_{\mathrm{min}} \gets \underset{i\in \{i \mid (\vb{h}_X)_i=1\}}{\operatorname{argmin}}~ \hat{p}_i$\;
  $\mathrm{Cut\_indexes} \gets \mathrm{Cut\_indexes} \cup \{i_{\mathrm{min}}\}$\;
}
$H_Z^{\mathrm{cut}} \gets \text{DeleteColumns}(H_Z, \mathrm{Cut\_indexes})$\;
$\hat{\vb{p}}^{\mathrm{cut}} \gets \text{DeleteEntries}(\hat{\vb{p}}, \mathrm{Cut\_indexes})$\;
$(\hat{\vb{e}}_X^{\mathrm{cut}}, \hat{\vb{p}}^{\mathrm{cut}}) \leftarrow \text{BP\_decode}(H_Z^{\mathrm{cut}}, \vb{s}_Z, \hat{\vb{p}}^{\mathrm{cut}}, T_{\mathrm{iter}})$\;
$\hat{\vb{e}}_X \gets \text{PadWithZeros}(\hat{\vb{e}}_X^{\mathrm{cut}}, \mathrm{Cut\_indexes})$\; 

\Return{$\hat{\vb{e}}_X$}
\end{algorithm}

\begin{algorithm}[t]
\caption{Alternative to line~8-11 in Algorithm~\ref{alg_1}}
\label{alg_2}
\setcounter{AlgoLine}{6}
\For{each $i$ in $\mathrm{Cut\_indexes}$}{
    \NoNumber
    $\hat{p}_i \gets 0$\;
}
$(\hat{\vb{e}}_X, \hat{\vb{p}}) \leftarrow \text{BP\_decode}(H_Z, \vb{s}_Z, \hat{\vb{p}}, T_{\mathrm{iter}})$
\end{algorithm}

As discussed in the previous section, the main reason for the failure of the BP decoder is the degeneracy inherent in qLDPC codes.
To address this problem, we introduce a post-processing algorithm for the BP decoder called \textit{degeneracy cutting (DC)}.
The main idea of DC is to resolve the degeneracy introduced by stabilizer generators by removing the variable node with the lowest error probability among those involved in each generator.
More precisely, for every $X$-type stabilizer generator, which is represented by a row vector $\vb{h}_X$ of the $X$-type parity-check matrix $H_{X}$, we identify the variable node with the smallest error probability among those indices $i$ satisfying $(\vb{h}_X)_i = 1$ and remove it from the Tanner graph.
If several variable nodes have the same lowest error probability, the nodes to be removed are selected at random.
By doing so, either of the degenerate errors $\vb{e}_X$ or $\vb{e}_X'$ satisfying $\vb{e}_X+\vb{e}_X'=\vb{h}_X$ becomes invalid if the variable node supporting $\vb{e}_X$ or $\vb{e}_X'$ is removed.
For instance, if the variable node associated with $\vb{e}_X'$ is removed, then $\vb{e}_X'$ cannot be a valid solution.
Running the BP decoder again on the modified graph is then more likely to converge to the correct error $\vb{e}_X$ (see Fig.~\ref{fig_degeneracy_cutting} (c)).
Since our post-processing effectively cuts loops formed by degeneracies arising from stabilizer generators, we call our method \textit{degeneracy cutting}.
Note that this cut removes only local degeneracies tied to each $X$-type generator; global degeneracies from logical operators stay untouched.
Although deleting variable nodes can in principle remove some solutions, our BP+DC+OSD results below indicate that, for the codes studied here, the graph modification typically preserves the solutions found by OSD.
Thus, DC breaks the harmful short loops while typically preserving syndrome consistency and overall decoding accuracy.
In Sec.~\ref{sec: Numerical analysis}, we demonstrate that our approach is reliable in practice.

\begin{figure*}[t]
    \begin{center}
        \includegraphics[width=\linewidth]{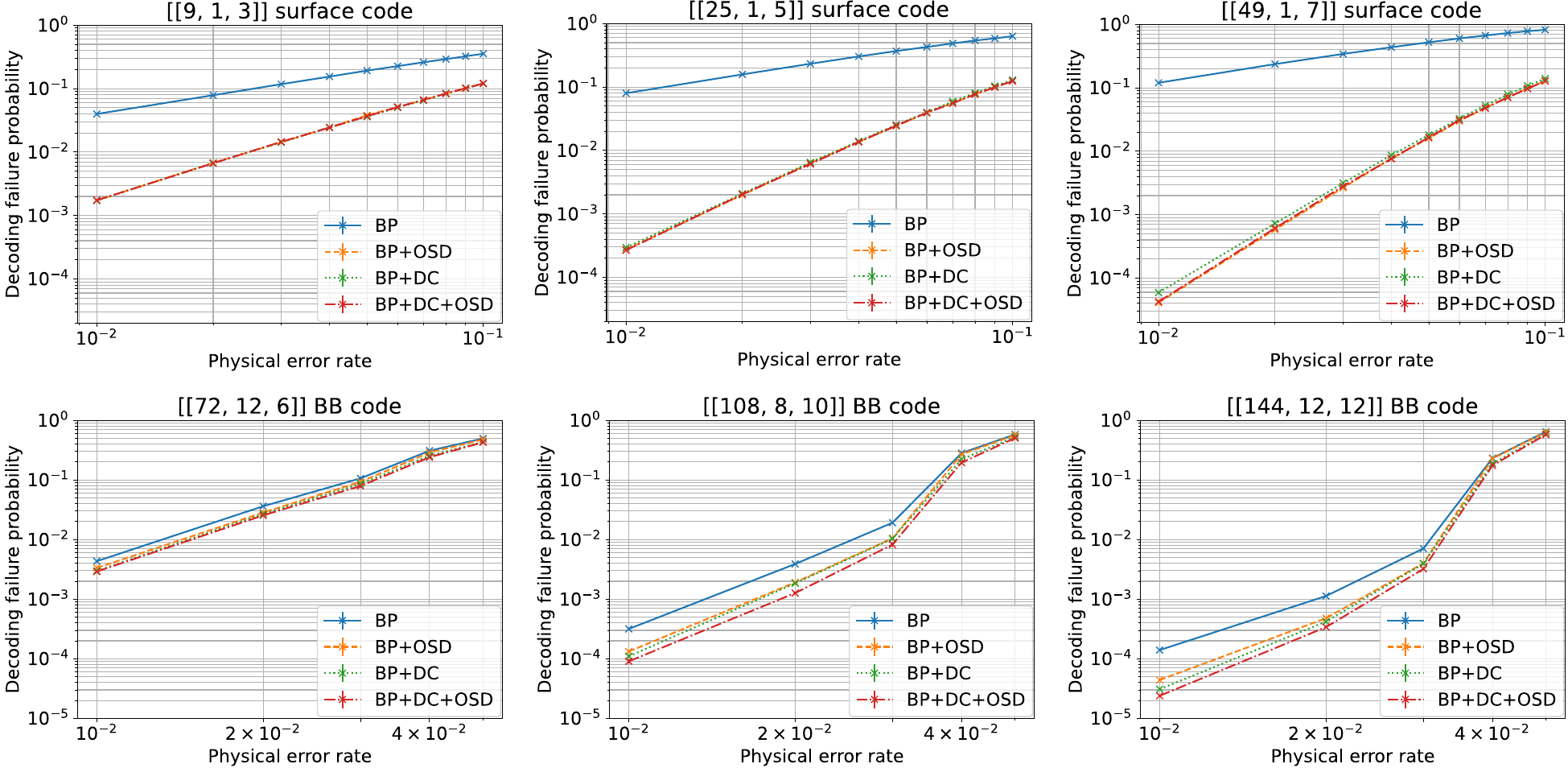}
        \caption{
        Performance of the BP decoder, BP+OSD decoder, BP+DC decoder, and BP+DC+OSD decoder for surface codes and BB codes under the code-capacity noise model.
        The x-axis indicates the physical error rate $p$, which represents the probability of a bit-flip on each physical qubit, while the y-axis shows the decoding failure probability, corresponding to the probability of decoding failure under different decoding strategies.
        The BP decoder is run up to $T_{\mathrm{iter}}=n$ iterations, where $n$ is the number of physical qubits, and we use the product-sum variant of BP for surface codes and the minimum-sum variant for BB codes.
        Error bars represent 95\% confidence intervals based on binomial statistics, but are smaller than the marker size and thus may not be visible in the plots. 
        While replacing the $O(n^3)$ computational cost of OSD step with the $O(n)$ computational cost of DC step, the BP+DC and BP+DC+OSD decoder achieves comparable performance to the BP+OSD decoder for surface codes.
        For BB codes, the BP+DC decoder achieves better performance than the BP+OSD decoder.
        }
        \label{fig_codecapacity}
    \end{center}
\end{figure*}

We summarize our decoding procedure in Algorithm~\ref{alg_1}.
In Algorithm~\ref{alg_1}, we use the following notation.
The routine $\text{BP\_decode}(H, \vb{s}, \vb{q}, T)$ denotes a single run of belief propagation on the Tanner graph of $H$ with syndrome $\vb{s}$, prior error probabilities $\vb{q}$, and at most $T$ iterations; it returns a hard-decision estimate together with the posterior error probabilities.
Given a qubit index set $S$, the operation $\text{DeleteColumns}(H, S)$ removes the columns of $H$ indexed by $S$, and $\text{DeleteEntries}(\vb{q}, S)$ removes the corresponding entries of $\vb{q}$, so that the reduced decoding problem is defined on the $n-|S|$ remaining variable nodes.
Conversely, $\text{PadWithZeros}(\hat{\vb{e}}, S)$ reinserts $0$ at every coordinate in $S$, mapping the reduced estimate back to the original $n$-qubit index set.
In this algorithm, each row vector $\vb{h}_X$ independently selects one variable node with the lowest estimated error probability within its support.
The cut set \texttt{Cut\_indexes} is then defined as the union of all selected candidates, and variable nodes indexed by the elements in \texttt{Cut\_indexes} are removed from the Tanner graph.
As a result, the procedure does not depend on the order in which rows are processed, and if multiple stabilizer generators nominate the same variable node, that node is removed only once.
In addition, we eliminate variable nodes by deleting the columns in \texttt{Cut\_indexes} from the $Z$-type parity-check matrix $H_Z$.
However, modifying the matrix directly is not necessary.
As shown in Algorithm~\ref{alg_2}, we can simply set the error probabilities of the eliminated nodes to 0.
This works because messages from variable nodes with probability 0 have no impact on the check nodes, and their probabilities remain 0 throughout the iterations.
In this sense, our method resembles guided decimation~\cite{yao2024belief, gong2024toward}, but unlike guided decimation, which sets error probabilities to 0 or 1 based on a global comparison of all variable nodes, our method fixes error probabilities to 0 based on local comparisons among variable nodes within each stabilizer generator.

We now discuss the computational cost of the BP+DC decoder.
The computational cost of an initial BP decoding is proportional to $n$, assuming the maximum number of iterations $T_{\mathrm{iter}}$ is fixed.
Since the row weight of the $X$-type parity-check matrix is bounded by $r$, identifying and eliminating the variable nodes with the lowest error probabilities across all rows of $H_X$ can also be done in $O(n)$ time.
Running the BP decoder again likewise has a computational cost proportional to $n$.
Therefore, the overall computational cost of the BP+DC decoder also scales linearly with $n$, which is comparable to that of the original BP decoder.

Moreover, the identification of variable nodes with the lowest error probabilities involves comparisons only among nodes within each $X$-type stabilizer generator, where the number of such nodes is upper bounded by a constant $r$. 
This property allows the elimination process to be implemented in parallel, much like the message-passing updates in the BP decoder.
This stands in stark contrast to other post-processing algorithms, which perform global comparisons across all variable nodes~\cite{du2022stabilizer, yao2024belief, gong2024toward, iolius2024closed, iolius2024almost, yin2024symbreak}.

In theory, repeating lines 4–10 in Algorithm~\ref{alg_1} could further improve the decoder’s accuracy.
However, we numerically observe that additional iterations do not enhance performance.
Thus, our algorithm modifies the Tanner graph only once, and we consider the decoder to be failed when the second BP outputs an estimate $\vb{\hat{e}}$ that does not match the syndrome.
As a result, unlike other post-processing algorithms that modify the Tanner graph multiple times~\cite{du2022stabilizer, yao2024belief, gong2024toward, yin2024symbreak}, the computational cost of BP+DC is at most double that of the original BP decoder, even in the worst case.

\subsection{Numerical analysis}
\label{sec: Numerical analysis}
To verify the effectiveness of the BP+DC decoder, we numerically calculate its decoding failure probabilities for surface codes and BB codes, and compare the results with those of other decoders in Fig.~\ref{fig_codecapacity}.
The simulations are performed under the \emph{code-capacity noise model}, in which
only data qubits suffer independent bit-flip errors with probability $p_i=p$, while stabilizer measurement is assumed to be noiseless.
Under this assumption, the prior error probabilities are represented as $\vb{p} = (p,\ldots,p)$.
Let us first discuss the results for surface codes.
From Fig.~\ref{fig_codecapacity}, we observe that the BP+DC decoder achieves performance approaching that of the BP+OSD decoder, while significantly reducing computational cost.
This indicates that the effect of degeneracy is effectively reduced by performing DC.
However, the performance of BP+DC is slightly worse than that of BP+OSD.
To confirm that this performance gap is not caused by an excessive elimination of variable nodes, we also perform OSD post-processing using data qubits that are not removed when BP+DC fails, which we refer to as BP+DC+OSD.
From Fig.~\ref{fig_codecapacity}, we see that the performance of BP+DC+OSD matches that of BP+OSD, suggesting that the graph modification introduced by DC does not eliminate the solutions found by OSD.

Next, we discuss the results for BB codes.
When applying BP+DC to BB codes, we find that accuracy improves if, in the second BP decoding step (line 9 of Algorithm~\ref{alg_1}), we replace the estimated error probabilities $\vb{\hat{p}}$ with the prior error probabilities $\vb{p}$ as input.
With this modification, Fig.~\ref{fig_codecapacity} shows that the BP+DC decoder can outperform the BP+OSD decoder, while maintaining a much lower computational cost.
We note that this modification is not always beneficial: for surface codes, we have observed that using the prior error probabilities $\vb{p}$ as input to the second BP step often degrades the performance of the decoder.
This is likely because DC effectively resolves the dominant local degeneracies for surface codes, making the informed posteriors from the first run a more valuable starting point than the original, unbiased prior. Conversely, for the BB codes, numerous short cycles remain after DC, making a reset to the original prior $\vb{p}$ a more robust strategy to prevent the second BP iteration from being dominated by, and thus amplifying, the biases present in the first run's posteriors.
While a formal analysis of this difference between BB codes and surface codes remains for future work, we provide additional numerical analysis in Appendices~\ref{app:overlap_bpdc_bposd} and~\ref{app:dc_flatness} to shed further light on the observed behavior of DC.

\section{Application beyond the code-capacity noise model}
\label{sec_beyond}
So far, we have focused on decoding errors under the code-capacity noise model, where syndrome measurements are assumed to be error-free and only data qubits experience errors.
However, in practical scenarios, errors may occur not only on data qubits but also on auxiliary qubits or during two-qubit gates used for syndrome measurements, making the decoding problem significantly more challenging.
In this section, we discuss how to apply the BP+DC decoder in such settings.
We first describe the detector error model~\cite{derks2024designing} and define decoding problems beyond the code-capacity noise model (Sec.~\ref{sec_beyond_1}).
Next, to capture the effect of degeneracy in the detector error model, we define a \textit{detector degeneracy matrix} and describe how to incorporate it into BP+DC decoding (Sec.~\ref{sec_beyond_2}).
Finally, we provide an explicit construction of the detector degeneracy matrix and evaluate the performance of our methods under phenomenological noise models (Sec.~\ref{sec_beyond_3}) and circuit-level noise models (Sec.~\ref{sec_beyond_4}).

\subsection{Detector error model}
\label{sec_beyond_1}
The effects of errors in syndrome extraction circuits can be captured by the \textit{detector error model}~\cite{derks2024designing,Higgott_2025,gidney2021stim}.
This model describes the relationship between physical errors and their impact on both \textit{detectors} and \textit{logical observables}.
A detector is a linear combination of measurement outcomes that will be $0$ in the absence of noise. 
Typically, a detector is defined as the parity of the measurement outcomes of a stabilizer generator over two successive rounds, which we will define more formally below.
A logical observable is a combination of measurement outcomes whose value corresponds to the result of measuring a logical Pauli operator.
The model is composed of three key elements: a \textit{detector check matrix}, which details which detectors are flipped by each error; a \textit{logical observable matrix}, which similarly tracks the impact on logical observables; and an associated noise model, which assigns a probability to the occurrence of each error.
By capturing the error behavior in this formalism, the detector error model allows a decoder to infer a likely error pattern given a measured syndrome, without needing details of the circuit and the specific noise models.

A \textit{location} is any single physical operation in the circuit, including a state preparation, a one- or two-qubit gate, an idling operation (i.e., $I$ gate), or a measurement.
We say a \textit{fault} occurs if a Pauli error is introduced at a location.
A fault at a single location may introduce one of several distinct Pauli errors.
For example, under the depolarizing channel, a fault at a single-qubit gate may introduce a Pauli $X, Y$ or $Z$ error, while a fault at a two-qubit gate can introduce one of the fifteen non-identity two-qubit Pauli errors.
We say that a Pauli error at a location flips a detector (or a logical observable) if its occurrence changes the outcome of the detector (or the logical observable).
It is often useful to group distinct Pauli errors that produce the exact same effect on detectors and logical observables, regardless of their type or the location where they occur. 
Such a class of equivalent errors is referred to as an \textit{error mechanism}.

Let $M$ be the number of detectors, $k$ be the number of logical observables, and $N$ be the number of error mechanisms.
In this work, we focus on decoding $X$-type errors using $Z$-type syndrome measurement results.
Let $m_{i,t}\in\mathbb{F}_2$ be the binary measurement outcome of the $i$-th $Z$-type stabilizer generator at measurement round $t$, where $i\in\{1,\ldots, m_Z\}$
and $t\in\{1,\ldots, T+1\}$.
We perform $T$ rounds of noisy measurements, followed by a final, noiseless measurement round at $t=T+1$.
A detector $d_{i,t}\in\mathbb{F}_2$ is represented as the difference (XOR) of consecutive outcomes,
\begin{equation}
    d_{i,t}
    = 
    \begin{cases*}
        m_{i,t} & if $t = 1$\\ 
        m_{i,t-1}\oplus m_{i,t} & if $t \geq 2$
    \end{cases*}.
\end{equation}
We define the \textit{error vector} as a row vector
\begin{equation}
    \vb{e}\in\mathbb{F}_2^N,
\end{equation}
where $\qty(\vb{e})_i=1$ if and only if the $i$-th error mechanism has occurred.
We define an independent error model as a set of $N$ independent error mechanisms.
Error mechanism $i$ occurs with a prior probability $p_i$, where each probability $p_i$ is calculated as the probability that an odd number of its constituent physical Pauli errors occur.
The set of these probabilities forms the prior vector
\begin{equation}
    \label{eq_priors}
    \mathrm{Pr}[(\vb{e}_i=1)]=p_i,\quad \vb{p}=\qty(p_1,\ldots,p_N).
\end{equation}

The action of all faults on all detectors is captured by the detector check matrix,
\begin{equation}
    H_{\mathrm{DCM}} \in \mathbb{F}_2^{M \times N},
\end{equation}
where $M=m_Z(T+1)$ is the number of detectors, and $\qty(H_{\mathrm{DCM}})_{ij}=1$ if and only if $j$-th error mechanism flips detector $i$.
In Tanner-graph point of view, $H_{\mathrm{DCM}}$ defines a bipartite graph whose variable nodes are error mechanisms and whose check nodes are detectors.
In addition, we define a \textit{logical operator matrix} as
\begin{equation}
    O\in\mathbb{F}_2^{k\times N},
\end{equation}
where the $(i,j)$ entry of $O$ is $1$ if and only if $j$-th error mechanism flips the logical-$Z$ measurement outcome of the $i$-th logical qubit.
Then, the decoding task in the detector error model can be formulated as follows: given a measured syndrome $\vb{s} =  \vb{e}H_{\mathrm{DCM}}^\top$, decoding estimates $\hat{\vb{e}}$ such that
\begin{align}
    &\hat{\vb{e}}H_{\mathrm{DCM}}^\top = \vb{s},\\
    &(\hat{\vb{e}}+\vb{e})O^\top = \vb{0}.
\end{align}

\subsection{Detector degeneracy matrix}
\label{sec_beyond_2}
Beyond the code-capacity noise model, degeneracy remains a key obstacle to decoding even under the detector-error model. 
Indeed, many low-weight errors $\mathbf e$ satisfy $\vb{e}H_{\mathrm{DCM}}^\top=\mathbf 0$ and thus escape the detection, while also leaving the logical information unchanged as $\vb{e}O^\top=\mathbf 0$.
Under the code-capacity noise model, such trivial errors are captured by the $X$-type parity-check matrix $H_X$.
Accordingly, the BP+DC decoder addresses the degeneracy by removing one variable node associated with each row of $H_X$.

To extend the BP+DC decoder beyond the code-capacity noise model, we introduce the \textit{detector degeneracy matrix} $H_{\mathrm{DDM}} \in \mathbb{F}_2^{M' \times N}$, where $M'$ is the number of low-weight trivial errors considered.
Each row $\vb{h}_{\mathrm{DDM}}$ of $H_{\mathrm{DDM}}$ represents a low-weight error that satisfies both $\vb{h}_{\mathrm{DDM}}H_{\mathrm{DCM}}^\top = \vb{0}$ and $\vb{h}_{\mathrm{DDM}}O^\top = \vb{0}$.
Consequently, the matrix $H_{\mathrm{DDM}}$ satisfies the condition $H_{\mathrm{DDM}}H_{\mathrm{DCM}}^{\top} = H_{\mathrm{DDM}}O^{\top} = 0$ and can be viewed as a generalization of the $X$-type parity-check matrix $H_X$ for bit-flip decoding under the code-capacity noise model.

The BP+DC decoder can then be generalized to more realistic noise models by replacing the parity-check matrices $H_Z$ and $H_X$ in Algorithm~\ref{alg_1} with the detector check matrix $H_{\mathrm{DCM}}$ and detector degeneracy matrix $H_{\mathrm{DDM}}$, respectively.
In Tanner-graph point of view, each row of $H_{\mathrm{DDM}}$ specifies a local subset of error-mechanism variable nodes that forms a trivial pattern, on which DC then acts by removing one low-posterior node before rerunning BP.
As prior error probabilities $\vb{p}$ in Algorithm~\ref{alg_1}, we use Eq.~\eqref{eq_priors}.
Importantly, this generalization can also be applied to other stabilizer-based post-processing methods~\cite{du2022stabilizer, yin2024symbreak}, which were previously limited to the code-capacity setting.
We note that, for post-processing methods that do not use the $X$-type parity-check matrix $H_{X}$, decoding can be performed by simply replacing the $Z$-type parity-check matrix $H_{Z}$ with the detector check matrix $H_{\mathrm{DCM}}$, without requiring the detector degeneracy matrix $H_{\mathrm{DDM}}$.
In the following, we present explicit constructions of $H_{\mathrm{DDM}}$ for the phenomenological and circuit-level noise models in BB codes.

Before presenting the explicit constructions, we introduce the notion of a local temporal window, which is useful for the circuit-level case of BB codes.
Because the syndrome-extraction circuit used in our simulations is repeated across measurement rounds, the local structure of trivial errors is invariant under shifts along the time direction.
We make this invariance explicit as follows.
Let $T_{\mathrm{loc}}$ be an integer (we will use $T_{\mathrm{loc}}=2$ or $3$).
We construct a reduced detector check matrix $H_{\mathrm{DCM}}^{(T_{\mathrm{loc}})}$ and the corresponding logical observable matrix $O^{(T_{\mathrm{loc}})}$ for only $T_{\mathrm{loc}}$ consecutive measurement rounds, numerically enumerate all low-weight error patterns $\vb{e}$ satisfying $\vb{e}{H_{\mathrm{DCM}}^{(T_{\mathrm{loc}})}}^{\top}=\vb{0}$ and $\vb{e}{O^{(T_{\mathrm{loc}})}}^{\top}=\vb{0}$ within this window, and then place each identified pattern at every possible starting round in the full $T$-round model.
We refer to this procedure as a \emph{local-window construction} of $H_{\mathrm{DDM}}$.
The local-window construction allows us to access weight-$w$ trivial patterns that span $T_{\mathrm{loc}}$ consecutive rounds without performing an exhaustive search over the full $T$-round detector model.

\subsection{Application to the phenomenological noise model}
\label{sec_beyond_3}
We begin by considering noise that acts on both data qubits and syndrome-measurement outcomes.
Specifically, we perform $T$ rounds of noisy $Z$-type syndrome measurements, where each measurement outcome is flipped with probability $p$, followed by a final noiseless round of syndrome measurement.
Independently, every data qubit experiences a bit-flip error with probability $p$ before each measurement round.
This noise model is referred to as the \textit{phenomenological noise model}.

The detector check matrix $H_{\mathrm{DCM}}$ for this model can be represented as
\begin{equation}
    \label{eq_DCM_phenomenological}
    \includegraphics[width=\linewidth]{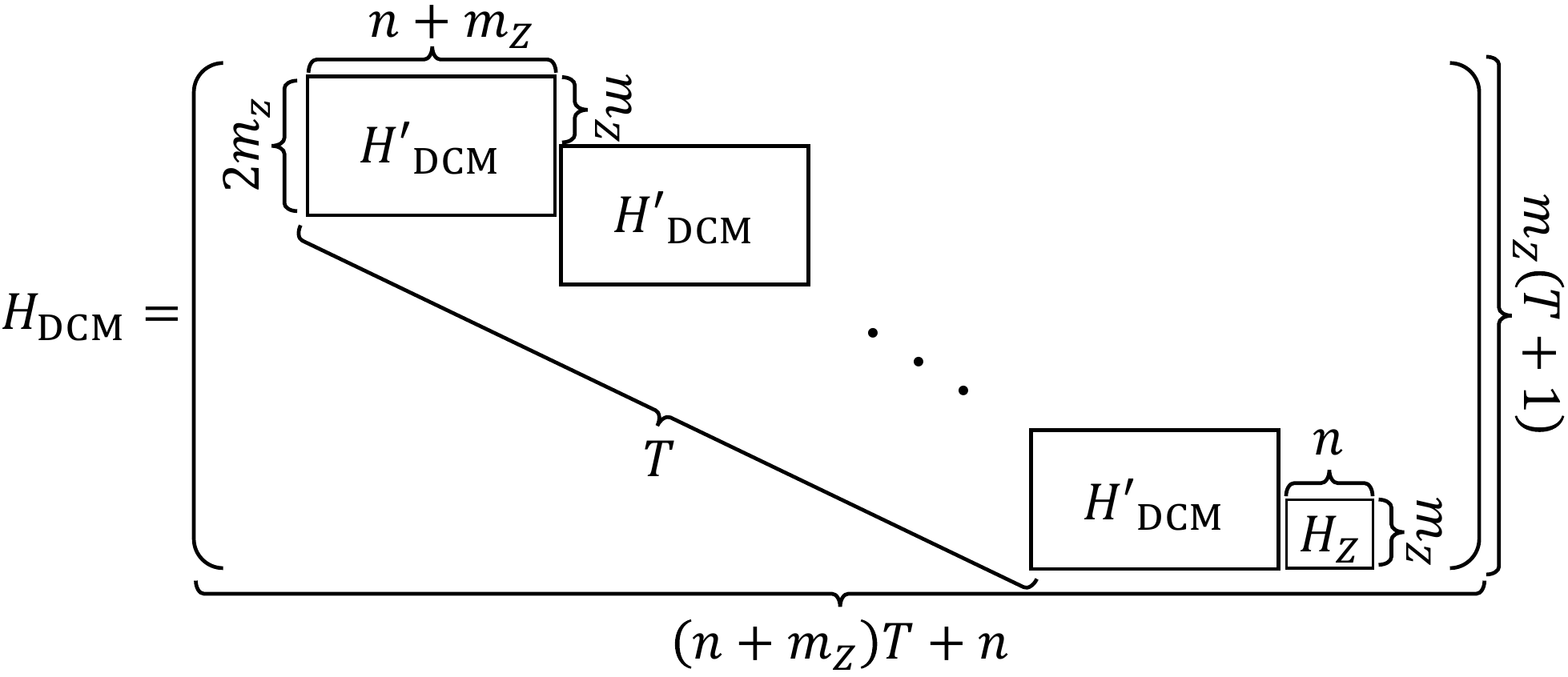}.
\end{equation}
This matrix is constructed by stacking $T$ identical submatrices $H_{\mathrm{DCM}}'\in\mathbb{F}_2^{2m_Z \times (n+m_Z)}$, each representing errors in the $t$-th round, and appending a single $H_Z\in\mathbb{F}_2^{m_Z\times n}$ block corresponding to data-qubit errors occurring just before the final noiseless measurement.
Here, $m_Z$ is the number of $Z$-type stabilizer generators and $n$ is the number of physical qubits of the code.
The submatrix $H_{\mathrm{DCM}}'$ is given by
\begin{equation}
H_{\mathrm{DCM}}' = 
\left(
\begin{array}{cc}
H_Z & I  \\ 
0 & I 
\end{array}
\right),
\end{equation}
where $I$ denotes the $m_Z\times m_Z$ identity matrix.
The first column block, $(H_Z^{\top}~0)^{\top}$, corresponds to bit-flip errors on the data qubits before the $t$-th syndrome measurement round.
Such errors affect all the subsequent syndromes, and thus flip the detector associated with the $t$-th round.
The second column block, $(I~I)^{\top}$, corresponds to measurement errors in the $t$-th round.
These errors flip the measurement outcome for the current round only, thereby flipping the detector for both the $t$-th and $(t+1)$-th rounds.
Note that the detector is defined as the XOR of the syndrome measurement outcomes from the previous and current rounds.
By this construction, the resulting detector check matrix $H_{\mathrm{DCM}}$ has row weight at most $r+2$ and column weight $c$.

\begin{figure}[t]
    \begin{center}
        \includegraphics[width=0.6\linewidth]{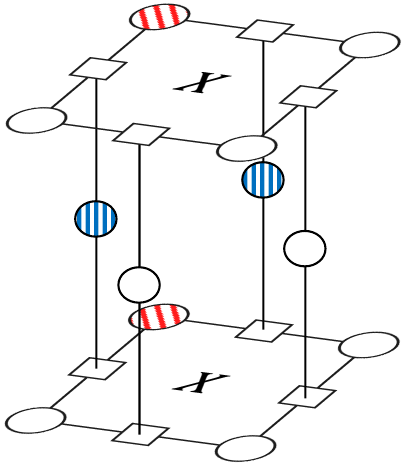}
        \caption{
        Schematic illustration of degeneracy arising from measurement errors in the phenomenological noise model.
        Circles and squares represent error mechanisms and detectors, respectively. Circles on the planar layers represent bit-flip errors on data qubits in the support of an $X$-type stabilizer generator; the lower and upper planes correspond to errors in the previous and current measurement rounds.
        Circles connecting the two planes represent measurement errors. Data-qubit errors on two consecutive rounds (circles with red diagonal patterns) are indistinguishable from measurement errors on the associated syndrome measurements (circles with blue vertical patterns).
        Therefore, if an error occurs on the data qubits, the BP decoder assigns the same error probabilities to all four error mechanisms, resulting in a decoding failure.
        }
        \label{fig_phenomenological}
    \end{center}
\end{figure}

\begin{figure*}[t]
    \begin{center}
        \includegraphics[width=\linewidth]{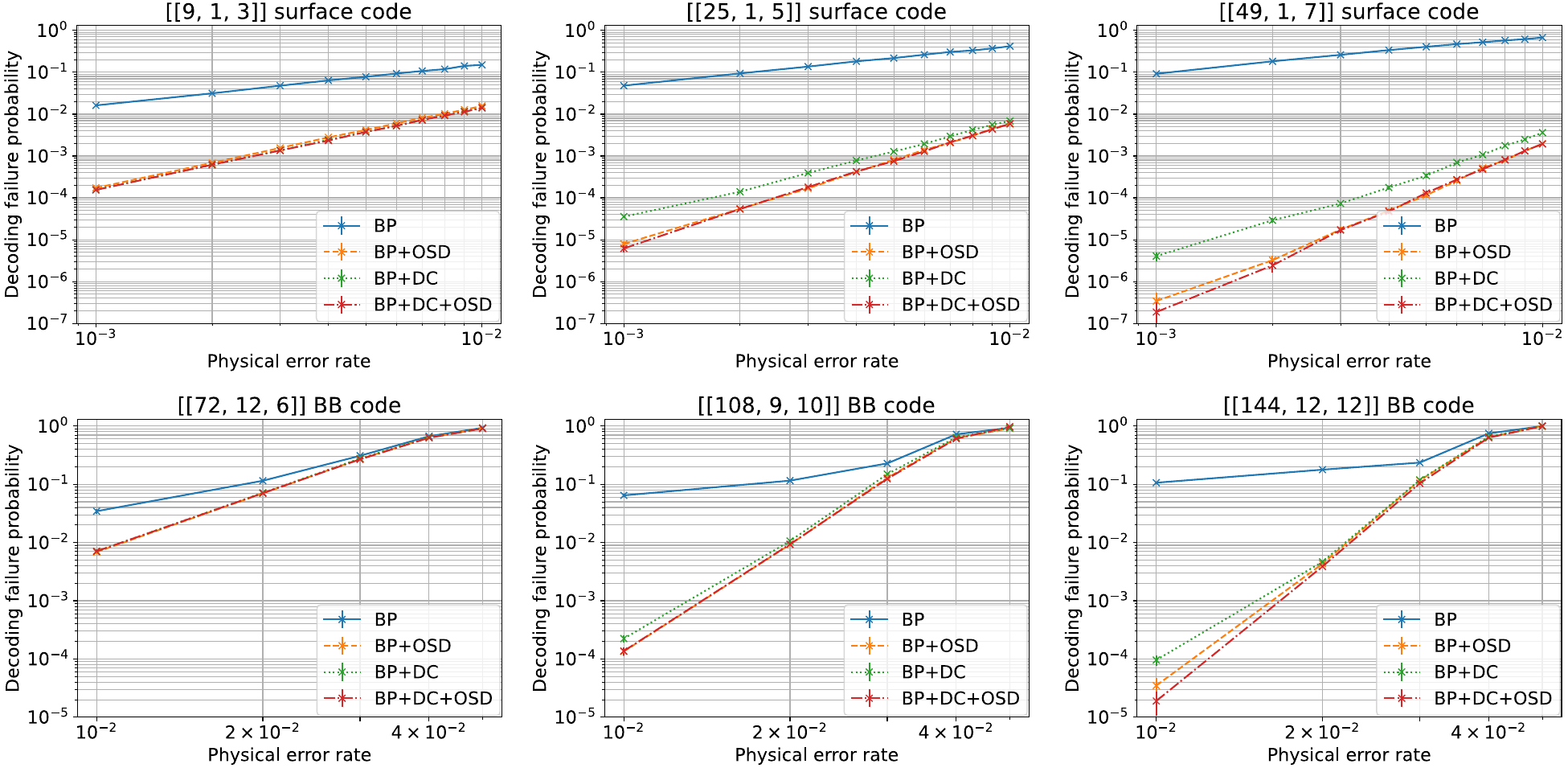}
        \caption{
        Performance of the BP decoder, BP+OSD decoder, BP+DC decoder, and BP+DC+OSD decoder for surface codes and BB codes under the phenomenological noise model.
        The x-axis indicates the physical error rate $p$, which represents the probability of an error on each physical qubit and syndrome measurements, while the y-axis shows the decoding failure probability, corresponding to the probability of decoding failure under different decoding strategies.
        The maximum number of BP iterations is set to $T_{\mathrm{iter}}=1000$ following the simulations performed in Ref.~\cite{iolius2024almost}, and we use the product-sum variant of BP for surface codes and the minimum-sum variant for BB codes.
        Error bars indicate the 95\% confidence intervals obtained by assuming binomial statistics for the number of decoding failures, but are smaller than the marker size and thus may not be visible in some plots.
        The BP+DC decoder achieves logical error suppression within an order of magnitude of the BP+OSD decoder while reducing computational cost from $O(N^3)$ to $O(N)$.
        }
        \label{fig_BB_phenomenological}
    \end{center}
\end{figure*}

Beyond the degeneracy observed under the code-capacity noise model, measurement errors introduce new types of degeneracy.
For example, consider a case in which a data qubit experiences errors at both the $t$-th and $(t+1)$-th rounds.
These errors affect the same detectors as measurement errors on the associated syndrome measurements at the $t$-th round (see Fig.~\ref{fig_phenomenological}).

To account for such degeneracy, we construct the detector degeneracy matrix $H_{\mathrm{DDM}}$ as
\begin{equation}
    \label{eq_DDM_phenomenological}
    \includegraphics[width=\linewidth]{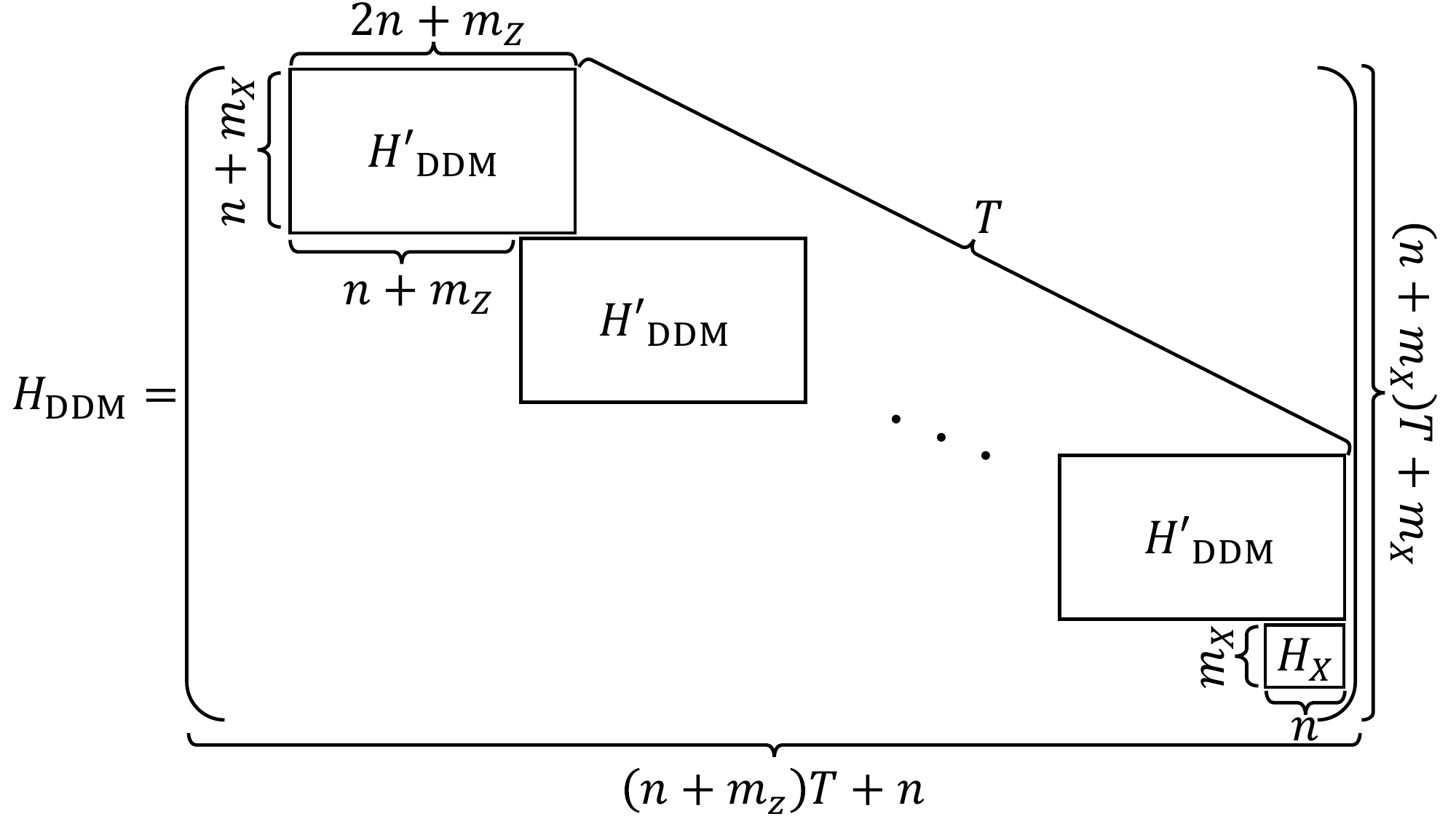}.
\end{equation}
This matrix is built by stacking $T$ submatrices $H_{\mathrm{DDM}}' \in \mathbb{F}_2^{(n + m_X) \times (2n + m_Z)}$, each representing degeneracy arising in the $t$-th round.
This is followed by a single $H_X$ block, corresponding to degeneracies caused by data-qubit errors prior to the final noiseless measurement round.
The submatrix $H_{\mathrm{DDM}}'$ is defined as
\begin{equation}
H_{\mathrm{DDM}}' = 
\left(
\begin{array}{ccc}
H_X & 0 & 0 \\ 
I & H_Z^{\top} & I
\end{array}
\right),
\end{equation}
where $I$ is the $n\times n$ identity matrix.
The three-column blocks correspond to data-qubit errors in the $t$-th round, measurement errors in the $t$-th round, and data-qubit errors in the $(t+1)$-th round, respectively.
The top row block $(H_X ~ 0 ~ 0)$ represents degeneracy from $X$-type stabilizer generators, while the bottom row block $(I ~ H_Z^{\top} ~ I)$ captures degeneracy due to measurement errors, as exemplified in Fig.~\ref{fig_phenomenological}.
Consequently, the resulting detector degeneracy matrix $H_{\mathrm{DDM}}$ satisfies $H_{\mathrm{DDM}} H_{\mathrm{DCM}}^{\top} = 0$ and has row and column weights upper-bounded by $\max\{r, c+2\}$.

By using the detector degeneracy matrix $H_{\mathrm{DDM}}$ constructed as in Eq.~\eqref{eq_DDM_phenomenological}, we apply the BP+DC decoder to surface codes and BB codes under the phenomenological noise model.
The numerical results are presented in Fig.~\ref{fig_BB_phenomenological}.
In our simulations, we set the total number of noisy syndrome measurement rounds to $T = d$.
We observe that the BP+DC decoder achieves performance comparable to that of the BP+OSD decoder, especially for the BB codes, while reducing the computational cost from $O(N^3)$ to $O(N)$.
This demonstrates the effectiveness of our decoder in handling the phenomenological noise model.

\subsection{Application to the circuit-level noise model}
\label{sec_beyond_4}
\begin{figure*}[t]
    \begin{center}
        \includegraphics[width=\linewidth]{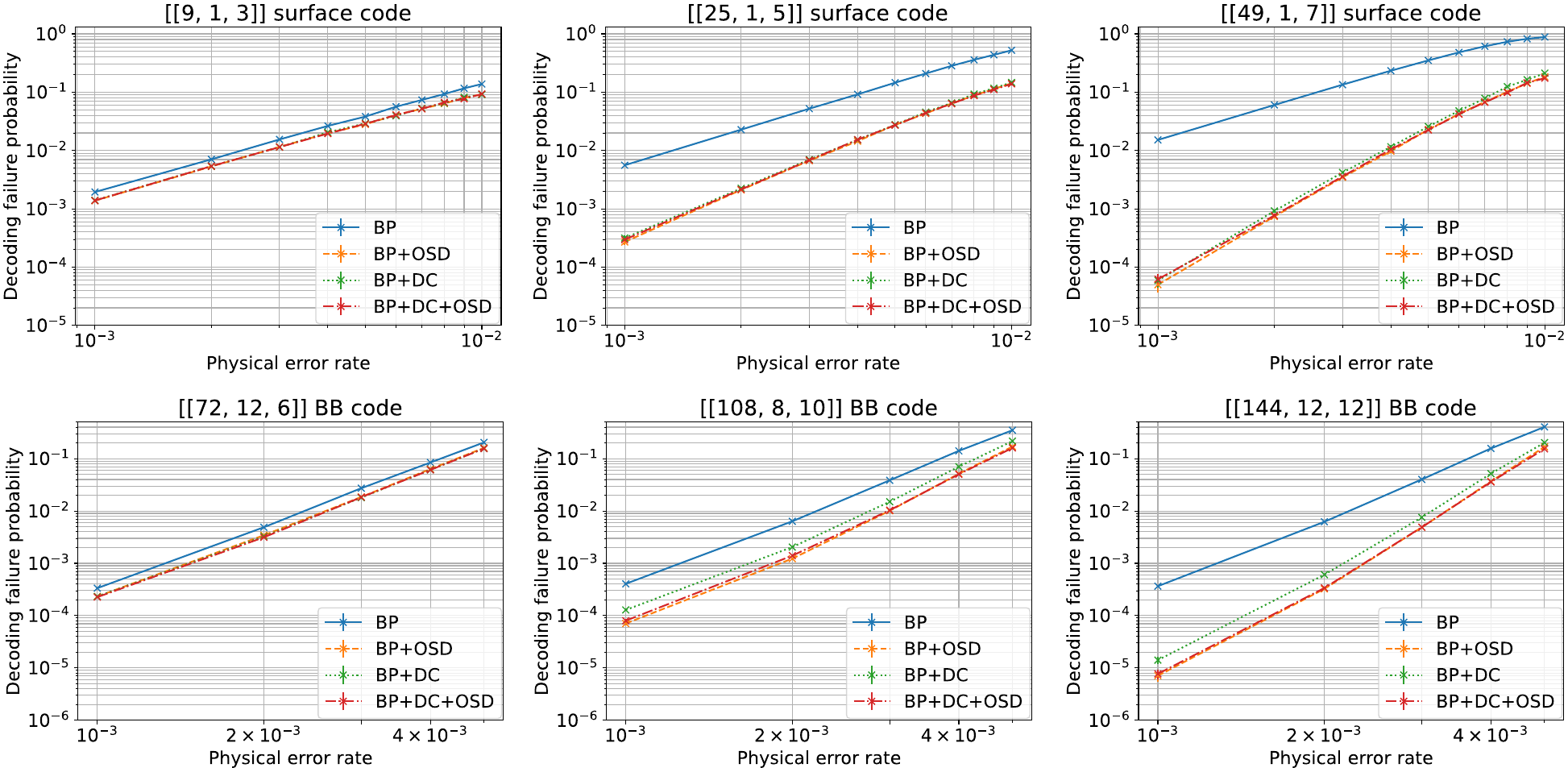}
        \caption{
        Performance of the BP decoder, BP+OSD decoder, BP+DC decoder, and BP+DC+OSD decoder for surface codes and BB codes under the circuit-level noise model.
        The x-axis indicates the physical error rate $p$, while the y-axis shows the decoding failure probability, corresponding to the probability of decoding failure under different decoding strategies.
        The maximum number of BP iterations is set to $T_{\mathrm{iter}}=1000$ following the simulations performed in Ref.~\cite{iolius2024almost}, and we use the product-sum variant of BP for surface codes and the minimum-sum variant for BB codes.
        Error bars indicate the 95\% confidence intervals obtained by assuming binomial statistics for the number of decoding failures, but are smaller than the marker size and thus may not be visible in some plots.
        The BP+DC decoder achieves logical error suppression within an order of magnitude of the BP+OSD decoder while reducing computational cost from $O(N^3)$ to $O(N)$.
        }
        \label{fig_BB_circuitlevel}
    \end{center}
\end{figure*}

Next, we apply the BP+DC decoder to the \textit{circuit-level noise model}.
In this model, state preparation errors, single- and two-qubit gate errors, idling errors, and measurement errors all occur with probability $p$ at each corresponding operation in the syndrome extraction circuit.
Specifically, single-qubit gates including idling gates are followed by $X$, $Y$, or $Z$ errors, each occurring with probability $p/3$; two-qubit gates are followed by one of the 15 non-identity two-qubit Pauli errors $\{P_1\otimes P_2\}_{P_1,P_2\in\{I,X,Y,Z\}}\setminus\{I\otimes I\}$, each with probability $p/15$; orthogonal states are prepared in state preparations with probability $p$; and measurement results are flipped with probability $p$.
The error probability $p$ is denoted as the physical error rate.

It is typical to assume that the noisy syndrome extraction circuit is repeated for $T$ rounds, followed by a single round of noiseless syndrome measurements.
For simplicity, we focus on decoding $X$-type errors using $Z$-type syndrome measurement results.
To construct the detector degeneracy matrix $H_{\mathrm{DDM}}$, we identify low-weight trivial error patterns beyond those captured by the stabilizer generators, namely error patterns $\vb{e}$ of weight at most $w$ satisfying $\vb{e}H_{\mathrm{DCM}}^\top = \vb{0}$ and $\vb{e}O^\top=\vb{0}$.
Such patterns typically combine data-qubit errors with measurement errors, or arise from errors propagating through two-qubit gates.
The rows of $H_{\mathrm{DDM}}$ are then formed by these identified trivial errors together with the original $X$-type stabilizer generators from $H_X$, and we set $w=5$ in our main simulations.
The explicit construction, which differs between surface codes and BB codes, is described in Appendix~\ref{app:ddm_weight_truncation}; the identification is a pre-computation and does not affect the runtime scaling of the BP+DC decoder itself.

Figure~\ref{fig_BB_circuitlevel} shows the performance of the BP+DC decoder for surface codes and BB codes under the circuit-level noise model.
In this simulation, we use the syndrome measurement schedule discussed in Ref.~\cite{o2024compare} for surface codes and in Ref.~\cite{bravyi2024high} for BB codes, which we review in Appendix~\ref{sec_circuitlevel_BB} for BB codes.
To perform the simulations, we calculate the error probabilities $\vb{p} = (p_1, \ldots, p_N)$ using the \texttt{stim} simulator~\cite{gidney2021stim}, based on the code provided in Ref.~\cite{gong2024toward}.
We set the total number of noisy syndrome measurement rounds to $T = d$ in the simulation.
Even in the circuit-level noise setting, the BP+DC decoder achieves logical error suppression comparable to that of the BP+OSD decoder, while reducing the computational cost from $O(N^3)$ to $O(N)$.
For the BB codes, a noticeable performance gap remains between BP+OSD and BP+DC.
Still, the fact that the performance of BP+DC+OSD is comparable to that of BP+OSD indicates that the graph modification introduced by DC does not eliminate the solutions found by OSD.
This is consistent with the interpretation that the gap originates from missing degeneracy in $H_{\mathrm{DDM}}$, rather than from BP+DC discarding solutions that BP+OSD exploits.
While a formal analysis of this dependence remains an important direction for future work, we provide additional numerical analyses in Appendix~\ref{app:ddm_weight_truncation} to show how the performance of BP+DC improves as the maximum weight $w$ of the trivial errors included in $H_{\mathrm{DDM}}$ is increased.

\section{Conclusion and Outlook}
\label{sec_conclusion}
In this work, we proposed degeneracy cutting (DC), a computationally efficient post-processing technique for the BP decoder that relies on decisions made within the support of individual stabilizer generators. 
Our method addresses decoding failures caused by degeneracy through the local elimination of variable nodes that contribute to low-weight degenerate errors.  
The resulting BP+DC decoder retains the favorable linear scaling in its computational cost and the inherent suitability for parallel implementation, while achieving a performance comparable to that of the BP+OSD decoder.

We demonstrated the effectiveness of the BP+DC decoder under the code-capacity noise model for both surface codes and BB codes.  
In particular, for BB codes, our method outperforms BP+OSD in both accuracy and computational cost.  
Furthermore, we extended the method to more realistic noise models---including phenomenological and circuit-level noise models---by introducing the detector degeneracy matrix, a generalization of the $X$-type parity-check matrix that captures degeneracy under such models.  
Our numerical results show that BP+DC achieves performance close to BP+OSD even in these more realistic and complex settings, while maintaining low computational cost.

Our results highlight the potential of locally defined and highly parallelizable decoding strategies for scalable quantum computing architectures.
A key advantage of the BP+DC decoder is its favorable computational scaling and high degree of parallelizability.
These characteristics are important for mitigating bottlenecks at the classical--quantum interface, where classical decoding circuits must be integrated with the qubits themselves.
A decoder such as BP+DC, which avoids global information exchange and is structured for parallel processing, can facilitate more hardware-friendly decoder implementations~\cite{sunami2025transversalsurfacecodegamepowered,PRXQuantum.6.010101,Barber_2025,Terhal_2015}.
By reducing the classical computation overhead associated with decoding, our method provides a lightweight solution that serves as a practical alternative to more computationally intensive decoders.

A promising direction for future research is improving performance by refining our methodology. 
For example, since DC can be executed in parallel and in linear time, it may also be applied intermittently during BP iterations with slight modifications to the algorithm. 
Instead of setting the error probabilities of selected variable nodes to $0$ as in Algorithm~\ref{alg_2}, these probabilities could be gradually reduced by a fixed factor at multiple points during BP iterations. 
Moreover, in decoding under a circuit-level noise model, we have constructed a detector degeneracy matrix based on trivial-error patterns of weight up to $w=5$, using the local-window construction for BB codes and exhaustive enumeration for surface codes.
The construction of the detector degeneracy matrix can be more flexible, and we expect that its performance can be further improved, e.g., by incorporating weight-6 local trivial errors.

Another promising direction is to extend our methodology to handle both X- and Z-type errors simultaneously. While this work focused on X-type errors, realistic noise models often involve correlated errors (e.g., Y-errors) that affect both bases. Addressing such correlations requires extending the detector error model to a unified framework~\cite{bravyi2024high}, which introduces additional low-weight degeneracies~\cite{beni2025tesseract}. While it is known that several decoders can handle the effect of those degeneracies~\cite{ott2025decision, beni2025tesseract, muller2025improved, maan2025decoding}, we anticipate that the BP+DC strategy can also be adapted by constructing a generalized detector degeneracy matrix that captures such effects.

Moreover, generalizing BP+DC to non-CSS qLDPC codes is also an interesting direction for future research.
Since our current method exploits the orthogonality between $X$-type and $Z$-type parity-check matrices, it cannot be directly applied to non-CSS codes.
Nevertheless, we expect that the core idea of degeneracy cutting—namely, the local identification and suppression of small-weight degeneracies—may remain useful in more general settings.

In addition, our decoder may be improved by combining our method with recently proposed modified BP techniques.
For instance, incorporating code automorphisms~\cite{koutsioumpas2025automorphism} or memory effects~\cite{muller2025improved} into the BP+DC framework may enhance decoding performance while preserving low computational cost and locality.  
Our detector degeneracy matrix could also be adapted based on the behavior of oscillating bits during BP iterations~\cite{wang2025fully}.  
We leave these explorations to future research.

A key theoretical question that remains open is whether a decoding threshold can be rigorously proven for BP and its post-processing variants like BP+OSD and BP+DC. While our numerical results show that BP+DC achieves a performance comparable to BP+OSD, a formal proof of a threshold for either decoder remains elusive. Establishing such a proof would not only clarify how post-processing techniques affect threshold behavior but would also contribute to a deeper understanding of the fundamental properties of qLDPC codes.

\section*{DECLARATION OF COMPETING INTERESTS.} H. Yamasaki and S. Tamiya are employees, and K. Tsubouchi is an intern at Nanofiber Quantum Technologies (NanoQT), Inc\@.

\section*{Data availability}
The data that support the findings of this article are not publicly available upon publication because it is not technically feasible and/or the cost of preparing, depositing, and hosting the data would be prohibitive within the terms of this research project. The data are available from the authors upon reasonable request.

\section*{Acknowledgements}
The authors acknowledge Akihisa Goban, Shinichi Sunami, and Yutaka Hirano for their feedback and support.

\bibliography{bib.bib}

\appendix
\section{Overlap analysis between BP+DC and BP+OSD on common Monte Carlo samples}
\label{app:overlap_bpdc_bposd}
\begin{table*}[t]
    \centering
    \caption{
    Overlap between BP+DC and BP+OSD on $100{,}000{,}000$ Monte Carlo samples under the code-capacity noise model at $p=0.01$. The column ``BP-step failed'' gives the number of sampled error-syndrome instances for which the first BP step is syndrome-inconsistent, so that post-processing is required for both BP+DC and BP+OSD. The remaining four columns show, within this subset, the numbers of instances for which both decoders succeed, both fail, only BP+DC succeeds, and only BP+OSD succeeds.
    }
    \label{tab:overlap_bpdc_bposd}
    \begin{tabular}{lccccc}
        \hline
        Code & BP-step failed &
        Both succeed &
        Both fail &
        BP+DC only &
        BP+OSD only \\
        \hline
        $[[49,1,7]]$ surface code
        & $11{,}995{,}211$
        & $11{,}990{,}292$
        & $3{,}098$
        & $11$
        & $1{,}810$ \\
        \hline
        $[[144,12,12]]$ BB code
        & $13{,}769$
        & $9{,}073$
        & $2{,}456$
        & $1{,}740$
        & $500$ \\
        \hline
    \end{tabular}
\end{table*}
To better understand the internal behavior of DC, we compare the success and failure events of BP+DC and BP+OSD on the \emph{same} Monte Carlo samples.
For each sampled error-syndrome instance, we classify the outcomes into four categories:
\begin{enumerate}
    \item both BP+DC and BP+OSD succeed,
    \item both fail,
    \item only BP+DC succeeds, and
    \item only BP+OSD succeeds.
\end{enumerate}

Table~\ref{tab:overlap_bpdc_bposd} summarizes the results for the code-capacity noise model of the $[[49,1,7]]$ surface code and the $[[144,12,12]]$ BB code at physical error rate $p=0.01$, each obtained from $10^8$ Monte Carlo samples.
For the surface code, BP-step fails in approximately $1.2 \times 10^{-1}$ of the samples.
Conditioned on this subset, BP+DC and BP+OSD agree on almost all instances.
Meanwhile, the rare disagreements overwhelmingly favor BP+OSD: among the $1{,}821$ disagreement events, $1{,}810$ are decoded correctly only by BP+OSD, while only $11$ are decoded correctly only by BP+DC.
Thus, for the surface code, BP+DC is best interpreted as a lightweight approximation to BP+OSD: it reproduces nearly all of the successful BP+OSD cases, while missing a very small subset of difficult instances.

The behavior is qualitatively different for the BB code.
In this case, DC is activated in only $1.4\times 10^{-4}$ of the samples, so the need for post-processing after the first BP pass is much rarer at this error rate.
However, on this hard subset, BP+DC and BP+OSD no longer behave as near-identical decoders.
Among the disagreement events, BP+DC succeeds in $1{,}740$ cases where BP+OSD fails, whereas BP+OSD succeeds in only $500$ cases where BP+DC fails.
Hence, for the BB code, BP+DC is not merely reproducing the same successful cases as BP+OSD, but is able to resolve a distinct subset of hard instances that BP+OSD misses.

These observations provide a more fine-grained interpretation of the numerical results in the main text.
For the rotated surface code, the small performance gap between BP+DC and BP+OSD comes from a very small number of hard instances on which OSD remains more powerful.
By contrast, for the BB code, the superior performance of BP+DC is associated with a different success set on the rare instances where DC is applied.
This overlap analysis therefore supports the view that the role of DC is code-family dependent: in some cases it acts mainly as a lightweight surrogate
for OSD, while in others it opens a distinct correction path that is not captured by OSD.

\section{Flatness of local posterior probabilities before applying DC}
\label{app:dc_flatness}

To gain further insight into how DC operates, we analyze the local structure of the posterior probabilities before applying DC.
Let $\vb{h}_X$ denote a row vector of $H_X$, and let $\hat{\vb{p}}$ be the posterior error probabilities after the first BP step.
To quantify how similar these local posterior probabilities are, we define
\begin{equation}
    \Delta(\vb{h}_X):=\max_{i\in \{i \mid (\vb{h}_X)_i=1\}} \hat{p}_i - \min_{i\in \{i \mid (\vb{h}_X)_i=1\}} \hat{p}_i.
\end{equation}
For shots where the first BP step fails, we classify each row vector $\vb{h}_X$ according to whether its local spread is below a threshold, namely whether
\begin{equation}
    \Delta(\vb{h}_X)\le \varepsilon_{\mathrm{gap}}
\end{equation}
holds for a threshold value $\varepsilon_{\mathrm{gap}}$.
A small spread means that the first BP step does not strongly distinguish among the candidate qubits within the support of $\vb{h}_X$, so that the qubit removed by DC is selected from a set of nearly equiprobable candidates. To avoid relying on a single threshold, Table~\ref{tab:dc_flatness} reports this fraction for several values $\varepsilon_{\mathrm{gap}}\in\{0.1,0.2,0.3,0.4,0.5\}$, for the $[[49,1,7]]$ surface code and the $[[144,12,12]]$ BB code under the code-capacity noise model at $p=0.01$. The results are conditioned on the first BP step failing and are measured over all evaluations of row vectors $\vb{h}_X$ on such failed shots.

\begin{table}[t]
    \centering
    \caption{Spread of the local posterior probability profiles used by DC under the code-capacity noise model at $p=0.01$. For each threshold $\varepsilon_{\mathrm{gap}}$, the entry gives the fraction of row vectors $\vb{h}_X$ whose local posterior spread satisfies $\Delta(\vb{h}_X)\le \varepsilon_{\mathrm{gap}}$. Results are conditioned on the first BP step failing, and percentages are rounded to one decimal place.}
    \label{tab:dc_flatness}
    \begin{tabular}{lccccc}
        \hline
        $\varepsilon_{\mathrm{gap}}$ & $0.1$ & $0.2$ & $0.3$ & $0.4$ &$0.5$ \\
        \hline
        $[[49,1,7]]$ surface code
        & $89.7\%$ & $90.1\%$ & $90.5\%$ & $91.0\%$ & $97.6\%$ \\
        $[[144,12,12]]$ BB code
        & $64.1\%$ & $66.7\%$ & $68.5\%$ &$69.9\%$ & $71.2\%$ \\
        \hline
    \end{tabular}
\end{table}

The two code families behave differently. For the surface code, the fraction is already about $89.7\%$ at the stringent threshold $\varepsilon_{\mathrm{gap}}=0.1$ and stays near $90\%$ up to $\varepsilon_{\mathrm{gap}}=0.3$, rising to $97.6\%$ at $\varepsilon_{\mathrm{gap}}=0.5$. Since the fraction remains high even for the smallest threshold, the local supports on which DC acts have genuinely flat posterior profiles for most supports, rather than profiles that are merely bounded in spread. For the BB code, the fraction increases more gradually, from $64.1\%$ at $\varepsilon_{\mathrm{gap}}=0.1$ to $71.2\%$ at $\varepsilon_{\mathrm{gap}}=0.5$. This smoother dependence reflects a broader distribution of local spreads, so that for the BB code DC acts on supports with bounded spread rather than on strictly flat profiles. These results also directly quantify how often the node removed by DC is selected from locally comparable candidates: for the surface code the large majority of selections occur on nearly flat supports essentially independent of the threshold, whereas for the BB code a substantial but threshold-dependent fraction does.

Taken together, these results show that the local posterior structure on which DC acts depends on the code family.
The local profiles examined by DC are markedly flatter for the surface code than for the BB code.
We emphasize that this diagnostic depends only on the spread $\Delta(\vb{h}_X)$ of the local posterior probabilities.
The purpose of this analysis is therefore to provide a simple quantitative characterization of how often DC acts on local supports whose posterior probabilities are mutually comparable.

\section{Circuit-level noise model for BB codes}
\label{sec_circuitlevel_BB}
In this appendix, we provide details of the circuit-level noise model used for BB codes.
Section~\ref{sec_measurementcircuit} introduces the syndrome measurement circuit proposed in Ref.~\cite{bravyi2024high}, along with the noise model considered in this work.
Then, we describe the resulting detector check matrix in Sec.~\ref{sec_DCM_circuitlevel}.

\subsection{Syndrome measurement circuit}
\label{sec_measurementcircuit}
In this section, we describe the syndrome measurement circuit for BB codes as introduced in Ref.~\cite{bravyi2024high}.
Consider a $2n$-qubit quantum circuit, where $n$ of the qubits are data qubits, $m_X = n/2$ qubits are auxiliary qubits for $X$-type syndrome measurements, and the remaining $m_Z = n/2$ qubits are auxiliary qubits for $Z$-type syndrome measurements.
We divide the data qubits into two groups, $L$ and $R$, corresponding to the left and right halves of the parity-check matrices $H_X = (A|B)$ and $H_Z = (B^{\top}|A^{\top})$.
The matrices $A=A_1+A_2+A_3$ and $B_1+B_2+B_3$ are defined in Sec.~\ref{sec_bb_code}.
We label the data qubits in $L$ and $R$, as well as the $X$- and $Z$-type auxiliary qubits, by $i = 1,\ldots, n/2$, and denote them as $q(L, i)$, $q(R, i)$, $q(X, i)$, and $q(Z, i)$, respectively.
For a permutation matrix $\Pi$, we write $j = \Pi(i)$ if the matrix $\Pi$ has a one in row $i$ and column $j$.

The circuit performs $T$ rounds of syndrome measurements using the following procedure.
We begin by initializing the $Z$-type auxiliary qubits:
\begin{enumerate}
    \setcounter{enumi}{-1}
    \item for $i=1$ to $n/2$ do
    \begin{itemize}
        \item $\texttt{Idle}(q(X, i))$
        \item $\texttt{InitZ}(q(Z, i))$
        \item $\texttt{Idle}(q(L, i))$
        \item $\texttt{Idle}(q(R, i))$
    \end{itemize}
\end{enumerate}
Here, \texttt{Idle} denotes the identity operation (i.e., doing nothing), and \texttt{InitZ} initializes the qubit to $\ket{0}$.
Then, the following eight steps are repeated for $T$ rounds:
\begin{enumerate}
    \item for $i=1$ to $n/2$ do
    \begin{itemize}
        \item $\texttt{InitX}(q(X, i))$
        \item $\texttt{CNOT}(q(R, A_1^{\top}(i)),~q(Z,i))$
        \item $\texttt{Idle}(q(L, i))$
    \end{itemize}
    \item for $i=1$ to $n/2$ do
    \begin{itemize}
        \item $\texttt{CNOT}(q(X,i),~q(L, A_2(i)))$
        \item $\texttt{CNOT}(q(R, A_3^{\top}(i)),~q(Z,i))$
    \end{itemize}
    \item for $i=1$ to $n/2$ do
    \begin{itemize}
        \item $\texttt{CNOT}(q(X,i),~q(R, B_2(i)))$
        \item $\texttt{CNOT}(q(L, B_1^{\top}(i)),~q(Z,i))$
    \end{itemize}
    \item for $i=1$ to $n/2$ do
    \begin{itemize}
        \item $\texttt{CNOT}(q(X,i),~q(R, B_1(i)))$
        \item $\texttt{CNOT}(q(L, B_2^{\top}(i)),~q(Z,i))$
    \end{itemize}
    \item for $i=1$ to $n/2$ do
    \begin{itemize}
        \item $\texttt{CNOT}(q(X,i),~q(R, B_3(i)))$
        \item $\texttt{CNOT}(q(L, B_3^{\top}(i)),~q(Z,i))$
    \end{itemize}
    \item for $i=1$ to $n/2$ do
    \begin{itemize}
        \item $\texttt{CNOT}(q(X,i),~q(L, A_1(i)))$
        \item $\texttt{CNOT}(q(R, A_2^{\top}(i)),~q(Z,i))$
    \end{itemize}
    \item for $i=1$ to $n/2$ do
    \begin{itemize}
        \item $\texttt{CNOT}(q(X,i),~q(L, A_3(i)))$
        \item $\texttt{MeasZ}(q(Z,i))$
        \item $\texttt{Idle}(q(R, i))$
    \end{itemize}
    \item for $i=1$ to $n/2$ do
    \begin{itemize}
        \item $\texttt{MeasX}(q(X,i))$
        \item $\texttt{InitZ}(q(Z,i))$
        \item $\texttt{Idle}(q(L, i))$
        \item $\texttt{Idle}(q(R, i))$
    \end{itemize}
\end{enumerate}
Here, \texttt{InitX} initializes the qubit to $\ket{+}$; \texttt{CNOT} applies a CNOT gate with the first qubit as control and the second as target; \texttt{MeasZ} and \texttt{MeasX} perform $Z$- and $X$-basis measurements, respectively.

In the circuit-level noise model, each operation described above is followed by a potential error characterized by a physical error rate $p$.
Specifically:
\begin{itemize}
    \item \texttt{Idle} is followed by $X$, $Y$, or $Z$ errors, each occurring with probability $p/3$.
    \item \texttt{CNOT} is followed by one of the 15 non-identity two-qubit Pauli errors $\{P_1\otimes P_2\}_{\{P_1,P_2\}\in\{I,X,Y,Z\}}\setminus\{I\otimes I\}$, each with probability $p/15$.
    \item \texttt{InitZ} and \texttt{InitX} prepare the incorrect states $\ket{1}$ and $\ket{-}$, respectively, with probability $p$.
    \item \texttt{MeasZ} and \texttt{MeasX} yield flipped outcomes with probability $p$.
\end{itemize}

\subsection{Detector check matrix}
\label{sec_DCM_circuitlevel}
Based on the syndrome measurement schedule described in Sec.~\ref{sec_measurementcircuit}, we construct the detector check matrix $H_{\mathrm{DCM}}$.
As noted in the main text, we focus on decoding $X$-type errors using $Z$-type syndrome measurement outcomes obtained from $\texttt{MeasZ}(q(Z,i))$, while discarding the $X$-type syndrome measurement results from $\texttt{MeasX}(q(X,i))$.
Furthermore, if multiple error events yield the same $Z$-type syndrome outcomes and the same logical effect, we treat them as a single aggregated error mechanism, whose probability is given by the total probability that an odd number of the constituent error events occur.
For example, Pauli errors $IX$, $ZX$, $IY$, and $ZY$ occurring after $\texttt{CNOT}(q(R, A_1^{\top}(i)),~q(Z,i))$ in step 1, and $XX$, $XY$, $YX$, and $YY$ occurring after $\texttt{CNOT}(q(R, A_3^{\top}(i)),~q(Z,i))$ in step 2, are grouped as a single effective error mechanism with total probability $8p/15+O(p^2)$.
This grouping, along with the resulting effective error probabilities, can be efficiently computed using the \texttt{stim} simulator~\cite{gidney2021stim}.

The detector check matrix $H_{\mathrm{DCM}}$ for this model can be represented as
\begin{equation}
    \label{eq_DCM_circuitlevel}
    \includegraphics[width=\linewidth]{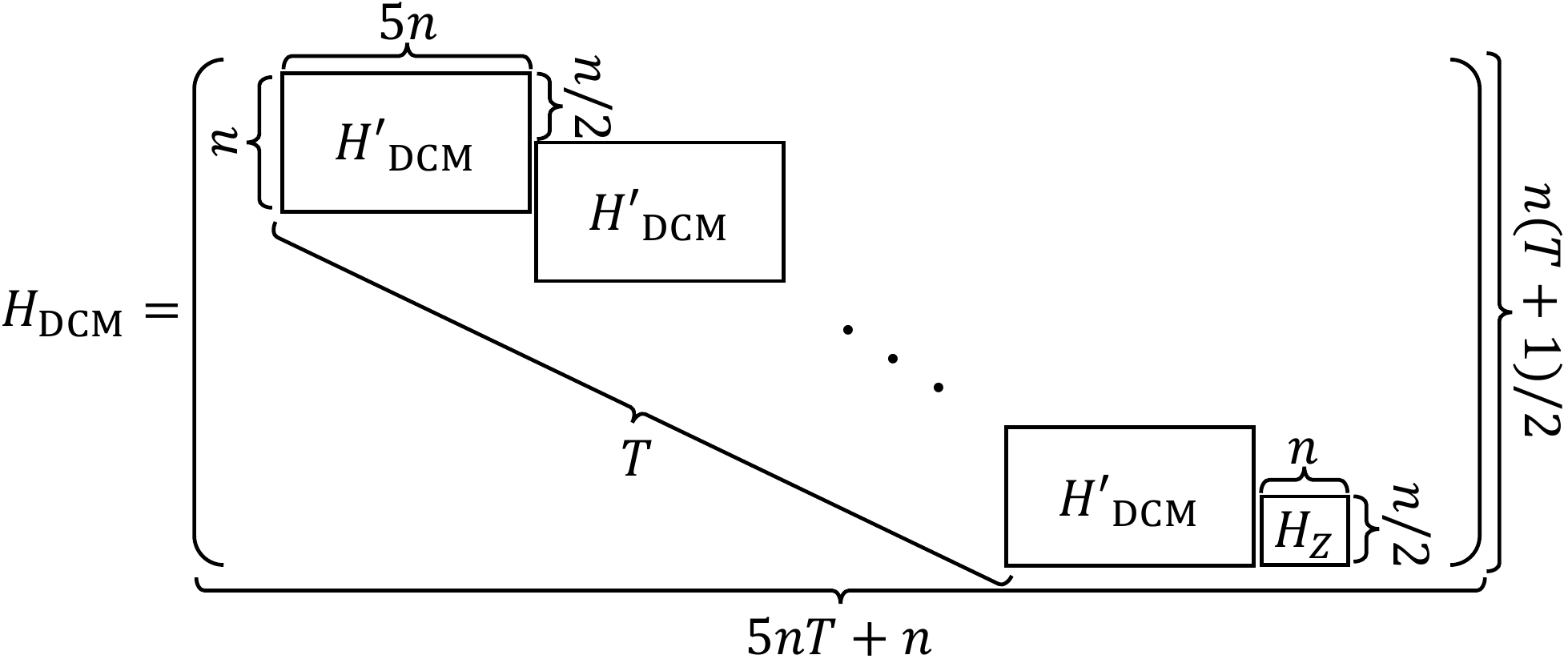}.
\end{equation}
This matrix is constructed by stacking $T$ submatrices $H_{\mathrm{DCM}}' \in \mathbb{F}_2^{n \times 5n}$, each of which represents the effect of errors occurring in the $t$-th round, followed by a single $H_Z$ block that corresponds to data-qubit errors occurring before the final noiseless measurement round (i.e., errors on $\texttt{Idle}(q(L, i))$ and $\texttt{Idle}(q(R, i))$ at step 8 in the $T$-th round).
Each submatrix $H_{\mathrm{DCM}}'$ is defined as
\begin{widetext}
\begin{equation}
H_{\mathrm{DCM}}' = 
\left(
\begin{array}{cc|c|cccc|ccc}
B^{\top} & A^{\top} & I & \bar{B}_1^{\top} & \bar{A}_1^{\top} & B_3^{\top}       & A_2^{\top}       & A_2^{\top}\bar{B}_2^{\top} & A_2^{\top} B_3^{\top}             & 0\\ 
0   & 0   & I & B_1^{\top}       & A_1^{\top}       & \bar{B}_3^{\top} & \bar{A}_2^{\top} & \bar{A}_2^{\top} B_2^{\top} & \bar{A}_2^{\top} \bar{B}_3^{\top} & \bar{A}_2^{\top} B^{\top}
\end{array}
\right),
\end{equation}
\end{widetext}
where each submatrix is an $n/2$-dimensional square matrix.  
Here, $I$ is an identity matrix, $\bar{A}_1 = A_2 + A_3$, $\bar{A}_2 = A_3 + A_1$, $\bar{A}_3 = A_1 + A_2$, $\bar{B}_1 = B_2 + B_3$, $\bar{B}_2 = B_3 + B_1$, and $\bar{B}_3 = B_1 + B_2$.

The first and second column blocks represent bit-flip errors on the data qubits after $\texttt{Idle}(q(L, i))$ and $\texttt{Idle}(q(R, i))$ during step 0 or step 8 in the previous round.
The third column block accounts for measurement errors from $\texttt{MeasZ}(q(Z,i))$ at step 7 in the $t$-th round.
The next four blocks correspond to data-qubit errors that occur during the $t$-th measurement round and affect detectors in both the $t$-th and $(t+1)$-th rounds.
Specifically, the fourth, fifth, sixth, and seventh blocks represent:
the bit-flip errors on $L$-type data qubits after $\texttt{CNOT}(q(L, B_1^{\top}(i)),~q(Z,i))$ at step 3;
on $R$-type data qubits after $\texttt{CNOT}(q(R, A_1^{\top}(i)),~q(Z,i))$ at step 1;
on $L$-type data qubits after $\texttt{CNOT}(q(L, B_2^{\top}(i)),~q(Z,i))$ at step 4;
and on $R$-type data qubits after $\texttt{CNOT}(q(R, A_3^{\top}(i)),~q(Z,i))$ at step 2, respectively.
The final three blocks capture bit-flip errors propagated from auxiliary qubits used for measuring $X$-type stabilizer generators.
In particular, the eighth, ninth, and tenth blocks correspond to bit-flip errors on $X$-type auxiliary qubits after $\texttt{CNOT}(q(X,i),~q(R, B_2(i)))$ at step 3,
$\texttt{CNOT}(q(X,i),~q(R, B_1(i)))$ at step 4,
and $\texttt{CNOT}(q(X,i),~q(R, B_3(i)))$ at step 5, respectively.
By this construction, the resulting detector check matrix $H_{\mathrm{DCM}}$ has row weight $35$ and column weight $6$.

\section{Dependence on the weight of trivial errors incorporated into the detector degeneracy matrix}
\label{app:ddm_weight_truncation}

In this appendix, we study how the decoding performance of the BP+DC decoder under the circuit-level noise model depends on the maximum weight $w$ of the trivial errors incorporated into the detector degeneracy matrix $H_{\mathrm{DDM}}$.
For each value of $w$, we construct $H_{\mathrm{DDM}}$ from the $X$-type stabilizer generators together with the numerically identified errors $\vb{e}$ of weight at most $w$ satisfying
\begin{equation}
    \vb{e}H_{\mathrm{DCM}}^\top=\vb{0},
    \qquad
    \vb{e}O^\top=\vb{0}.
\end{equation}
For surface codes, we enumerate them exhaustively over the full $T$-round detector model.
For BB codes, we exploit the time-translation symmetry of the syndrome extraction schedule, in which the circuit is identical across rounds: we define a {local temporal window} of $T_{\mathrm{loc}}$ consecutive rounds, enumerate the trivial-error patterns of weight at most $w$ supported within this window, and translate each identified pattern along the time direction to every admissible starting round of the full $T$-round model.
A window of $T_{\mathrm{loc}}$ rounds admits $T-T_{\mathrm{loc}}+1$ such placements.
For the codes and weights considered here, every relevant trivial-error pattern has temporal support contained within $T_{\mathrm{loc}}=3$ consecutive rounds, so we fix $T_{\mathrm{loc}}=3$ and obtain the full $H_{\mathrm{DDM}}$ from $T-2$ placements.
For example, a single placement suffices when $T=3$, whereas for $T=4$ two placements, covering rounds $1$--$3$ and $2$--$4$, reproduce the entire matrix.
The BP+DC results shown in Fig.~\ref{fig_BB_circuitlevel} correspond to the case $w=5$.

As an example, for BB codes we explicitly present the detector degeneracy matrix $H_{\mathrm{DDM}}$ for the case $w=3$ below.
Corresponding to the detector check matrix $H_{\mathrm{DCM}}$ described in Sec.~\ref{sec_DCM_circuitlevel}, the detector degeneracy matrix $H_{\mathrm{DDM}}$ for $w=3$ is constructed as
\begin{equation}
    \label{eq_DDM_circuitlevel}
    \includegraphics[width=\linewidth]{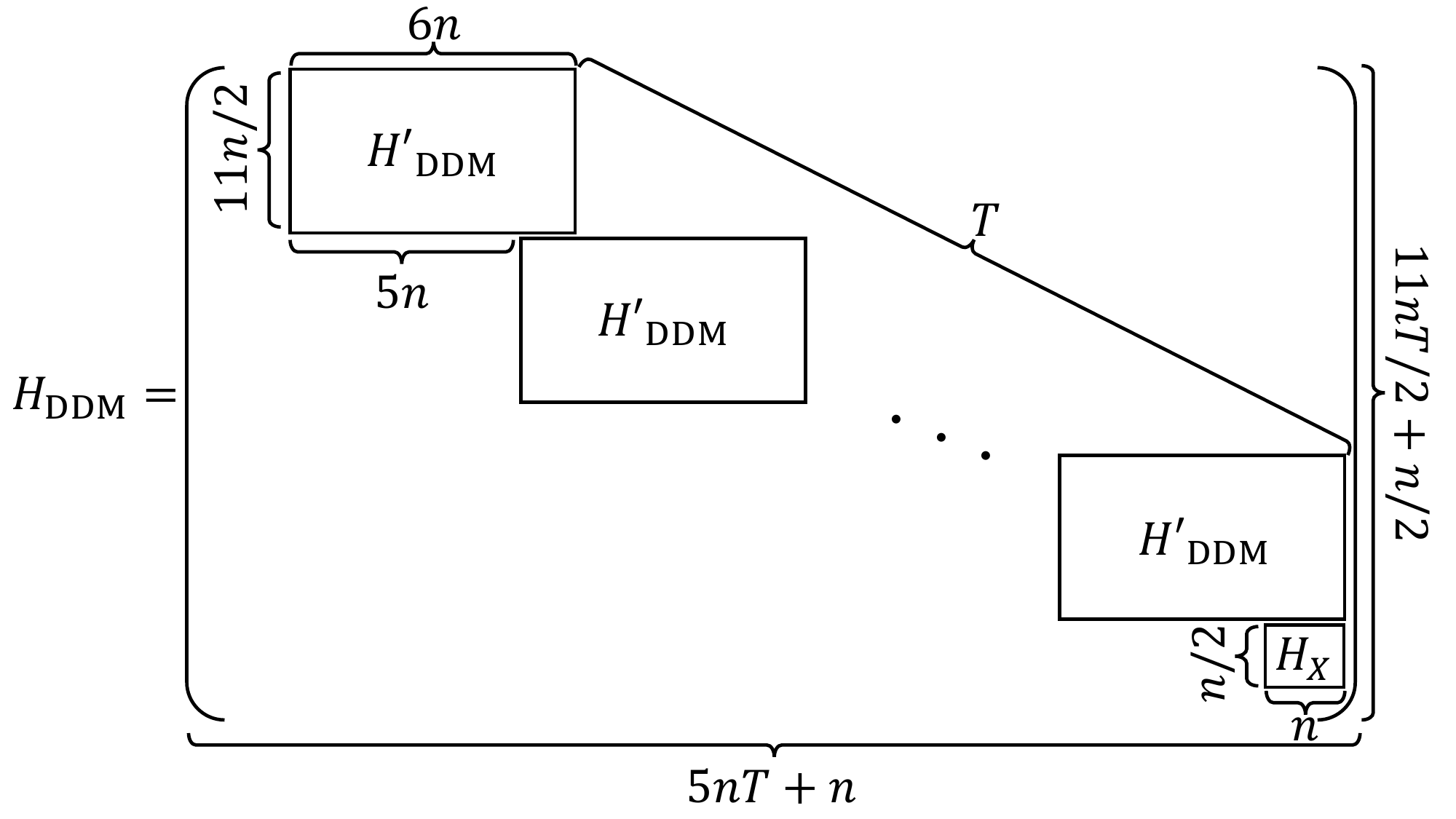}.
\end{equation}
The matrix is formed by stacking $T$ submatrices $H_{\mathrm{DDM}}' \in \mathbb{F}_2^{11n/2 \times 6n}$, each representing degeneracy introduced by errors in the $t$-th round.  
This is followed by a single $H_X$ block, corresponding to degeneracies due to data-qubit errors before the final noiseless measurement round.  
Each submatrix $H_{\mathrm{DDM}}'$ is defined as
\begin{equation}
H_{\mathrm{DDM}}' = 
\left(
\begin{array}{cc|c|cccc|ccc|cc}
A   & B & 0   & 0 & 0 & 0 & 0   & 0 & 0 & 0 & 0 & 0 \\ \hline
I   & 0 & B_1 & I & 0 & 0 & 0   & 0 & 0 & 0 & 0 & 0 \\
0   & I & A_1 & 0 & I & 0 & 0   & 0 & 0 & 0 & 0 & 0 \\
0   & 0 & B_2 & I & 0 & I & 0   & 0 & 0 & 0 & 0 & 0 \\
0   & 0 & A_3 & 0 & I & 0 & I   & 0 & 0 & 0 & 0 & 0 \\
0   & 0 & B_3 & 0 & 0 & I & 0   & 0 & 0 & 0 & I & 0 \\
0   & 0 & A_2 & 0 & 0 & 0 & I   & 0 & 0 & 0 & 0 & I \\ \hline
A_2 & 0 & 0   & 0 & 0 & 0 & B_2 & I & 0 & 0 & 0 & 0 \\
0   & 0 & 0   & 0 & 0 & 0 & B_1 & I & I & 0 & 0 & 0 \\
0   & 0 & 0   & 0 & 0 & 0 & B_3 & 0 & I & I & 0 & 0 \\
0   & 0 & 0   & 0 & 0 & 0 & 0   & 0 & 0 & I & \bar{A}_2 & 0
\end{array}
\right),
\end{equation}
where each submatrix is an $n/2 \times n/2$ matrix.  
The first two column blocks represent data-qubit errors before the $t$-th syndrome measurement round.  
The third column block corresponds to measurement errors on the $t$-th round.  
The next four blocks account for data-qubit errors during the $t$-th round.  
The subsequent three blocks correspond to errors propagated from auxiliary qubits.  
The final two blocks represent data-qubit errors before the $(t+1)$-th syndrome measurement round.

The row blocks of this matrix are structured to capture the distinct physical origins of the weight-$3$ trivial errors identified numerically. 
Specifically, we group the identified trivial error patterns into three main categories: (1) degeneracies from the original $X$-type stabilizer generators, (2) those arising from measurement errors interacting with data qubit errors across time steps, and (3) those caused by errors propagating from X-ancilla qubits to data qubits via CNOT gates.
The top row block captures type (1) degeneracies, with row weight $6$.
Note that $H_X = (A \mid B)$ is the $X$-type parity-check matrix.  
The next six row blocks describe degeneracies of type (2), each with row weight $3$.  
The final four row blocks capture degeneracies due to type (3), also with row weight $3$.  
By this construction, the resulting detector degeneracy matrix $H_{\mathrm{DDM}}$ satisfies $H_{\mathrm{DDM}} H_{\mathrm{DCM}}^{\top} = 0$ and $H_{\mathrm{DDM}} O^{\top} = 0$, and has row weight $6$ and column weight $7$.

We numerically verify that all weight-3 errors $\vb{e}$ satisfying $\vb{e}H_{\mathrm{DCM}}^\top = \vb{0}$ and $\vb{e}O^\top = \vb{0}$ are included as row vectors of $H_{\mathrm{DDM}}$ for all BB codes presented in Ref.~\cite{bravyi2024high}, except for the $[[108, 8, 10]]$ BB code.
For the $[[108, 8, 10]]$ BB code, we find that the following additional row block needs to be appended to each $H_{\mathrm{DDM}}'$ to ensure inclusion of all weight-3 trivial errors:
\begin{equation}
\left(
\begin{array}{cc|c|cccc|ccc|cc}
0   & 0 & 0   & 0 & 0 & 0 & 0   & 0 & 0 & A_1^2 A_3 + A_1 A_3^2 + I & 0 & 0
\end{array}
\right).
\end{equation}
In the numerical simulations for $w=3$, we used this modified detector degeneracy matrix $H_{\mathrm{DDM}}$ for the $[[108, 8, 10]]$ BB code.
Detector degeneracy matrix $H_{\mathrm{DDM}}$ for larger $w$ can also be constructed by a numerical search with $O(N^{\lceil w/2\rceil})$ time.
We note that this search can be carried out as a pre-computation and thus does not affect the runtime scaling of the BP+DC decoder itself.

\begin{figure*}[t]
    \centering
    \includegraphics[width=\textwidth]{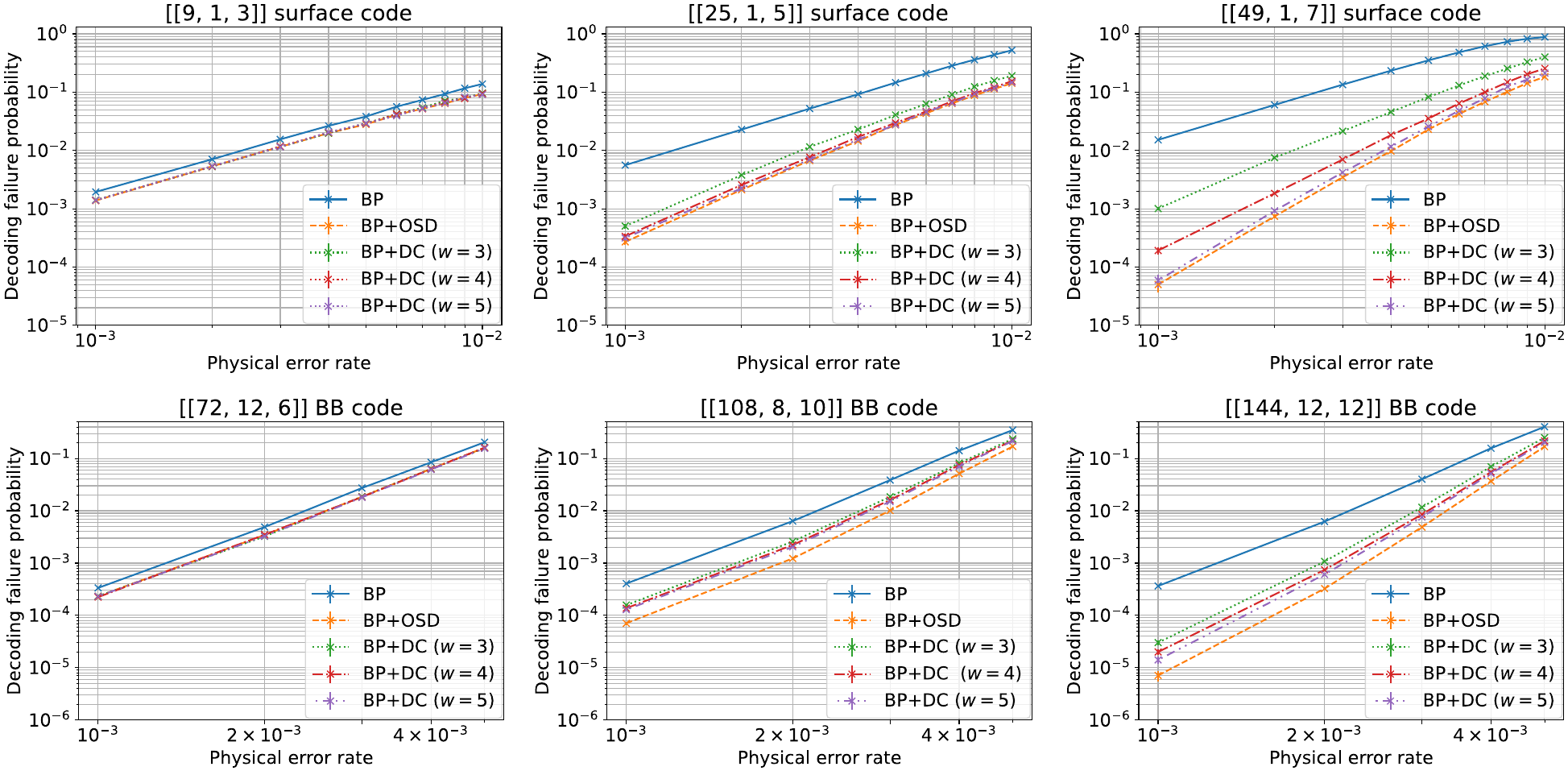}
    \caption{
    Dependence of the BP+DC decoder on the maximum weight $w$ of the trivial errors incorporated into the detector degeneracy matrix $H_{\mathrm{DDM}}$ under the circuit-level noise model.
    The curves labeled BP+DC $(w=3)$, BP+DC $(w=4)$, and BP+DC $(w=5)$ correspond to detector degeneracy matrices $H_{\mathrm{DDM}}$ constructed using trivial errors of weight at most $3$, $4$, and $5$, respectively, in addition to the original $X$-type stabilizer generators (see Appendix~\ref{app:ddm_weight_truncation} for the precise construction).
    The x-axis indicates the physical error rate $p$, while the y-axis shows the decoding failure probability, corresponding to the probability of decoding failure under different decoding strategies.
    The maximum number of BP iterations is set to $T_{\mathrm{iter}}=1000$ following the simulations performed in Ref.~\cite{iolius2024almost}, and we use the product-sum variant of BP for surface codes and the minimum-sum variant for BB codes.
    Error bars indicate the 95\% confidence intervals obtained by assuming binomial statistics for the number of decoding failures, but are smaller than the marker size and thus may not be visible in some plots.
    }
    \label{fig:circuitlevel_weight}
\end{figure*}

Figure~\ref{fig:circuitlevel_weight} compares the decoding failure probabilities obtained with $w=3$, $w=4$, and $w=5$.
For all codes considered here, increasing $w$ improves or preserves the performance of BP+DC and brings it closer to BP+OSD.
The improvement is especially pronounced for the larger codes.
In particular, for $[[144,12,12]]$ BB code, modifying $H_{\mathrm{DDM}}$ from $w=3$ to $w=5$ halves the decoding failure probability at $p=10^{-3}$, and for $[[49,1,7]]$ surface code, there is more than an order-of-magnitude improvement.
These results support the interpretation that a substantial part of the remaining performance gap between BP+OSD and BP+DC arises from degeneracies that are not captured by the smaller detector degeneracy matrices.
These results suggest that further enriching the detector degeneracy matrix may yield additional improvements.
We leave a systematic study of this dependence as an interesting direction for future work.

\end{document}